\documentclass[
aps,
floatfix,
letterpaper,
prx,
singlecolumn,
reprint,
superscriptaddress]{revtex4-2}
\usepackage[up,bf,raggedright]{titlesec}
\titleformat{\section}
  {\normalfont\fontsize{10}{12}\bfseries}{\thesection}{1em}{}
\titleformat{\subsection}
  {\normalfont\fontsize{10}{12}\bfseries}{\thesubsection}{1em}{}

\usepackage[sectionbib]{bibunits}
\defaultbibliographystyle{apsrev4-2} 
\defaultbibliography{Correlated_sensing} 

\usepackage[colorlinks=true, linkcolor=blue]{hyperref}% add hypertext capabilities

\usepackage{lipsum}
\usepackage{graphicx} 
\usepackage{amsfonts}
\usepackage{amsmath}
\usepackage{braket}
\usepackage{gensymb}
\usepackage{hyperref}
\usepackage{dsfont}
\usepackage[capitalize]{cleveref}
\usepackage{nicematrix}

\usepackage[utf8]{inputenc}
\usepackage[T1]{fontenc}

\newcommand{\nx}{n_1}
\newcommand{\ny}{n_2}

\newcommand{\rideal}{r_\text{ideal}}

\newcommand{\seq}{\mathord{=}}

\newcommand{\phiGx}{\phi_{C_a}}
\newcommand{\phiGy}{\phi_{C_b}}
\newcommand{\phiGxy}{\phi_{C_{a,b}}}

\newcommand{\phiL}{\phi_{L}}

\newcommand{\e}{\textrm{e}}

\newcommand{\cosp}[1][]{\cos\left(#1\right)}
\newcommand{\sinp}[1][]{\sin\left(#1\right)}

\date{\today}

\begin{document}
\begin{bibunit}[apsrev4-2]

\begin{abstract}
    Nitrogen vacancy (NV) centers in diamond are widely deployed as local magnetic sensors, using coherent, single qubit control to measure both time-averaged fields and noise with nanoscale spatial resolution. Moving beyond single qubits to multi-qubit control enables new sensing modalities such as measuring nonlocal spatiotemporal correlators, or using entangled states to improve measurement sensitivity. Here, we describe protocols to use optically unresolved NV center pairs and nuclear spins as multi-qubit sensors for measuring correlated noise, enabling covariance magnetometry at nanometer length scales. For NV centers that are optically unresolved but have spectrally resolved spin transitions, we implement a phase-cycling protocol that disambiguates magnetic correlations from variance fluctuations by alternating the relative spin orientations of the two NV centers. For NV centers that are both optically and spectrally unresolved, we leverage the presence of a third qubit, a $^{13}$C nucleus that is strongly coupled to one of the NV centers, to effect coherent single-NV spin flips and enable a similar phase-cycling protocol. For length scales around 10 nm, we create maximally entangled Bell states through dipole-dipole coupling between two NV centers, and use these entangled states to directly read out the magnetic field correlation, rather than reconstructing it from independent measurements of unentangled NV centers. Importantly, this changes the scaling of sensitivity with readout noise from quadratic to linear. For conventional off-resonant readout of the NV center spin state (for which the readout noise is roughly 30 times the quantum projection limit), this results in a dramatic sensitivity improvement. Finally, we demonstrate methods for the detection of high spatial- and temporal-resolution correlators with pairs of strongly interacting NV centers. 
\end{abstract}
\title{Multi-qubit nanoscale sensing with entanglement as a resource}
\author{Jared Rovny} 
\affiliation{Princeton University, Department of Electrical and Computer Engineering, Princeton, NJ 08544, USA}
\author{Shimon Kolkowitz}
\affiliation{University of California at Berkeley, Department of Physics, Berkeley CA 94720, USA}
\author{Nathalie P.\ de Leon}
\thanks{Corresponding author. Email: npdeleon@princeton.edu}
\affiliation{Princeton University, Department of Electrical and Computer Engineering, Princeton, NJ 08544, USA}

\maketitle

The development of multi-qubit sensors is a central goal in quantum sensing, with important applications in metrology \cite{Eldredge2018,Song2024} and the study of many body dynamics in quantum materials \cite{Rovny2024a}, where they can not only reveal previously hidden phenomena like spatiotemporal correlations \cite{Szankowski2015,Rovny2022} and non-Gaussian statistics \cite{Norris2016}, but also promise more sensitive measurements by using entanglement in the sensor as a resource \cite{Bollinger1996}. Nitrogen vacancy (NV) centers in diamond are robust nanoscale sensors that allow for coherent quantum control, enabling local measurements of quantities beyond average magnetic fields, and they have been successfully applied to nanoscale nuclear magnetic resonance \cite{Du2024,Allert2022}, and numerous condensed matter systems such as the study of magnetic phase transitions \cite{Dwyer2021,Machado2023a,Ziffer2024,Li2024,Xue2024}, driven magnonic excitations \cite{Zhou2021a,Simon2022a}, ballistic and diffusive conductivity \cite{Kolkowitz2015,Ariyaratne2018a}, and transport in graphene \cite{Andersen2019}. This combination of coherent control and target proximity makes NV centers an ideal platform for the development of multi-qubit sensors. 

The recent development of the covariance magnetometry protocol \cite{Rovny2022} introduced measurements of classical spatiotemporal magnetic field correlations using multiple NV centers, in which $\braket{B(r_a,t_a)B(r_b,t_b)}$ is directly measured using NV centers at distinct locations $r_\text{a,b}$ and times $t_\text{a,b}$. While the locations $r_\text{a,b}$ can be determined within a few nanometers of precision, the smallest lengthscale accessible to covariance magnetometry using two optically resolved NV centers is the optical diffraction limit, $|r_\text{a}-r_\text{b}| \gtrsim 250\,$nm, limiting their applicability as a general multi-qubit sensor.

The nanometer lengthscale regime is important in many contexts, for instance studying noise in thin-film superconductors \cite{Curtis2024} or spin excitations in magnetic insulators \cite{Chatterjee2019}, where correlation lengths are often between 1\,nm and 1\,$\mu$m. One method for accessing correlations at nanometer lengthscales is to operate at cryogenic temperatures, where the NV centers become spectrally resolvable in their optical transitions, enabling independent resonant optical readout; this method has recently been used to measure correlated charge fluctuations \cite{Ji2024a,Delord2024a}. Nevertheless, many systems of interest are restricted to higher temperatures (like hydrodynamic electron transport \cite{Lucas2018}, dynamics in high-$T_c$ cuprates \cite{Liu2025}, and biological systems \cite{Allert2022}), motivating the development of a more general multi-qubit protocol. This is challenging because the inability to optically resolve two NV centers requires their correlation to be inferred from photon statistics alone, where the low NV center readout fidelity and presence of systematic fluctuations can obscure the signatures of spin-spin correlations. 

Here we introduce super-resolution covariance magnetometry, which enables the measurement of two-point correlators below the diffraction limit in multiple contexts, and realizes NV center clusters as more versatile multi-qubit sensors. We first demonstrate a protocol to measure two-point correlators using two optically unresolved NV centers with resolvable spin transitions arising from their different crystallographic orientations; these resolved transitions allow independent spin manipulation, enabling a phase-cycling protocol that disambiguates magnetic correlations from variance fluctuations, related to a recently reported method using two NV centers in a single scanning diamond tip \cite{Huxter2024}. We then demonstrate a method for measuring correlations using two co-aligned NV centers that are both optically and spectrally unresolved, making use of nearby $^{13}$C nuclear spins to effect a $^{13}$C-mediated single-NV spin flip, once again enabling a phase cycling protocol despite the lack of independent control otherwise available. Finally, we describe multi-qubit sensing with strongly interacting NV center pairs that are closely-spaced ($\sim 10\,$nm), where NV-NV coupling becomes significant and covariance sensing with Bell pairs allows two-point correlators to be accessed in a much more direct way. We demonstrate entanglement-based protocols for sensing both spatial and temporal correlations, showing that entanglement-based covariance magnetometry offers much higher sensitivity than can be achieved using non-interacting NV centers in the low readout fidelity regime. 

\section{Detecting correlations between optically unresolved NV centers}

In covariance magnetometry \cite{Rovny2022} with two optically resolvable NV centers NVa and NVb, the Pearson correlation is computed between the distinct signals $S_a$ from NVa and $S_b$ from NVb. Here, we instead measure two optically unresolved NV centers (``sub-diffraction NV centers''), which appear in a confocal microscope as a single bright spot (\cref{fig:overview}A). While super-resolution techniques like charge state depletion (CSD) \cite{Chen2015} and deterministic emitter-switch microscopy (DESM) \cite{Chen2013} allow us to determine the locations of the two NV centers to calibrate their separation (\cref{fig:overview}A, right), these imaging techniques rely on collecting photons from a single NV center at a time, and therefore cannot be readily applied to covariance magnetometry. 

At the end of the covariance measurement we have access only to the total signal from both NV centers $S=S_a+S_b$, where the signal $S=\{s_i\}$ is a list containing the total number of photons $s_i$ collected from both NV centers in each experiment, and $i = 1...N$ indexes the $N$ total experiments. The basic method for accessing the two-NV correlation from this total signal relies on sensitive measurements of the variance $\text{Var}(S)$ of the joint distribution, which contains contributions from the individual NV center variances as well as an excess variance from correlations \cref{fig:overview}C. The key challenge in extracting the magnetic correlation from $\text{Var}(S)$ is that technical sources of excess variance arise not only from correlated environmental noise, such as correlated fluctuations in laser power and optical alignment, but also from technical fluctuations in the individual NV center statistics, which are now inaccessible because the NV centers are not optically resolvable.

To isolate the magnetic covariance from the measured total variance, we perform a four-step phase cycling experiment inspired by related methods in NMR \cite{Braunschweiler1983}, and similar to the method that was recently employed in \cite{Huxter2024}, in which we alternately prepare correlated and anticorrelated states by alternating the starting (or ending) spin orientation of each NV center in turn (\cref{fig:overview}D). These four experiment cycles, which we label $\mathcal{A}$ through $\mathcal{D}$, result in four expressions for the total measured photon variance:
\begin{align}
  \sigma^2_{S_\mathcal{A}} &= +2 \sigma_{a\uparrow}\sigma_{b\uparrow} r  + \sigma_{a\uparrow}^2 + \sigma_{b\uparrow}^2 \nonumber \\
    \sigma^2_{S_\mathcal{B}} &= -2 \sigma_{a\uparrow}\sigma_{b\downarrow} r  + \sigma_{a\uparrow}^2 + \sigma_{b\downarrow}^2 \nonumber\\
    \sigma^2_{S_\mathcal{C}} &= -2 \sigma_{a\downarrow}\sigma_{b\uparrow} r  + \sigma_{a\downarrow}^2 + \sigma_{b\uparrow}^2 \nonumber\\
    \sigma^2_{S_\mathcal{D}} &= +2 \sigma_{a\downarrow}\sigma_{b\downarrow} r  + \sigma_{a\downarrow}^2 + \sigma_{b\downarrow}^2,
\end{align}
where the inclusion of $\uparrow,\downarrow$ in the individual NV center variances accounts for possible deviations caused by the pulse protocol used to change the NV center's spin orientation. We then combine these results such that the correlation terms add constructively to find the covariance signal
\begin{align}
    \text{Cov}(S_a,S_b) \equiv \frac{1}{8}
    \left(\sigma^2_{S_\mathcal{A}} - \sigma^2_{S_\mathcal{B}} - \sigma^2_{S_\mathcal{C}} + \sigma^2_{S_\mathcal{D}}\right). \label{eq:Cov}
\end{align}
We use the mean value of the phase-cycled signal (which should be zero in an unbiased noise measurement) to report on any residual errors in the measured covariance and perform baseline subtraction \cite{Supp}.

We first demonstrate this protocol using two shallow NV centers with different orientations, such that the $m_s=0 \rightarrow m_s=-1$ transition of the NV center that is aligned with the magnetic field is detuned by 50 MHz from that of the misaligned NV center (\cref{fig:overview}B), allowing us to independently address each NV center's spin transition and implement the phase-cycling technique shown in \cref{fig:overview}D. We then use an external radiofrequency (RF) coil to apply a global, random phase $2\,$MHz AC signal, which we detect using a Hahn echo sequence for each NV center. Finally, we read out the spin states of the NV centers using spin-to-charge conversion (SCC) \cite{Shields2015} to reduce the readout noise, an important consideration for covariance magnetometry, where the sensitivity scales quadratically with readout noise \cite{Pfender2019,Rovny2022}. As we increase the strength of the RF signal, we observe an increase in covariance, which then oscillates as expected for a pure AC tone \cite{Rovny2022} and decreases after the NV centers accumulate a phase of $\pi$ (\cref{fig:overview}E). The sign of the covariance is negative because the two NV centers are sensitive to different projections of the external RF field; when we instead address the opposite spin transition of one NV center, the covariance changes sign as expected. For perfectly unbiased noise measurements, \cref{eq:Cov} predicts a maximum covariance amplitude $\text{Cov}(S_a,S_b) = \sigma_a\sigma_b r \approx 0.3 $, in good agreement with our results \cite{Supp}.

\section{Detecting correlations between optically and spectrally unresolved NV centers}

The method described above can be used for NV centers with resolved spin transitions due to their differing crystallographic alignments. However, this method measures covariance between different projections of the magnetic field, and furthermore, precludes the application of high magnetic fields (beyond approximately 100\,G) which can result in significantly degraded performance for NV centers misaligned with the applied field \cite{Epstein2005,Lai2009,Tetienne2012}. To accomplish phase cycling for optically unresolved, co-aligned NV centers with spectrally unresolved spin transitions, we make use of a $^{13}$C nuclear spin that is strongly coupled to one of the NV centers to selectively flip that NV center, as shown in \cref{fig:highfield13C}. Given the natural abundance (1.1\%) of $^{13}$C nuclear spins, we expect a single coupled $^{13}$C with a hyperfine coupling strength within a coherence time of $50\,\mu$s for about 90\% of NV center pairs \cite{Takacs2024}. This coherence time is the average observed for shallow NV centers under dynamical decoupling \cite{Sangtawesin2019}, and these statistics are consistent with our random sampling of 6 NV center pairs \cite{Supp}.

We use an XY8 dynamical decoupling sequence \cite{Gullion1990} to perform spectroscopy of the local $^{13}$C nuclear spin environment in the vicinity of two closely spaced NV centers (\cref{fig:highfield13C}A). When we sweep the interpulse spacing, we identify one incoherent spectral feature corresponding to the bath of $^{13}$C spins with weak NV center coupling, and a series of coherent features corresponding to individual $^{13}$C spins with stronger NV center coupling.  Having thus identified the resonance frequency of a strongly-coupled $^{13}$C spin (\cref{fig:highfield13C}B), we choose the interpulse spacing that brings the $^{13}$C-coupled NV center into resonance with this spin, and sweep the number of applied pulses to perform a coherent $^{13}$C-mediated spin flip on this NV center; this coherent spin flip will result regardless of the orientation of the $^{13}$C spin \cite{Kolkowitz2012,Taminiau2012,Zhao2012,Supp}. 

In order to make the four phase cycles as similar as possible for background cancellation during co-addition, we insert the extra gate pulses at the beginning of all four cycles; for cycles where the selective $^{13}$C-mediated spin flip is required, we choose an interpulse spacing $\tau_1$ that brings the NV center into resonance with the $^{13}$C, and for cycles where no selective spin flip is required, we instead use a similar interpulse spacing $\tau_0$ that is not resonant (\cref{fig:highfield13C}B,C). The selective spin flip on the $^{13}$C-coupled NV center, combined with the alternation of the final global pulse, allows us to accomplish the four-phase cycle; applying a global RF pulse as before, we are able to observe a covariance signal from two NV centers that are neither optically nor spectrally resolved (\cref{fig:highfield13C}D).

\section{Entanglement as a resource for super-resolution covariance magnetometry}

The super-resolution covariance magnetometry techniques described above use two noninteracting NV centers as a multi-qubit sensor. However, because the states of these NV centers are intrinsically uncorrelated, detecting correlated signals requires extracting a correlation from the product of two independent readout events with uncorrelated readout noise. This multiplicative step incurs a quadratic readout noise penalty \cite{Rovny2022}, making correlation measurements especially challenging when the readout noise is high, as is the case for room-temperature NV center sensing. NV centers located within $\sim$\,10 nm of each other open a new opportunity to use entanglement as a resource when their dipole-dipole coupling becomes strong enough to enable the creation of entangled states that are intrinsically correlated \cite{Gaebel2006,Neumann2010,Dolde2013,Dolde2014,Lee2023,Joas2024}. These entangled states can be used to encode the presence of a correlation directly in the state populations, effectively moving the comparative step from the post-readout multiplication event into the metrological event, removing the quadratic readout noise penalty.
We now show that using entanglement as a resource significantly improves the sensitivity of covariance magnetometry, and removes the need for SCC readout or variance-based detection.

\begin{figure*}[ht]
	\centering
	\includegraphics[width=0.99\textwidth]{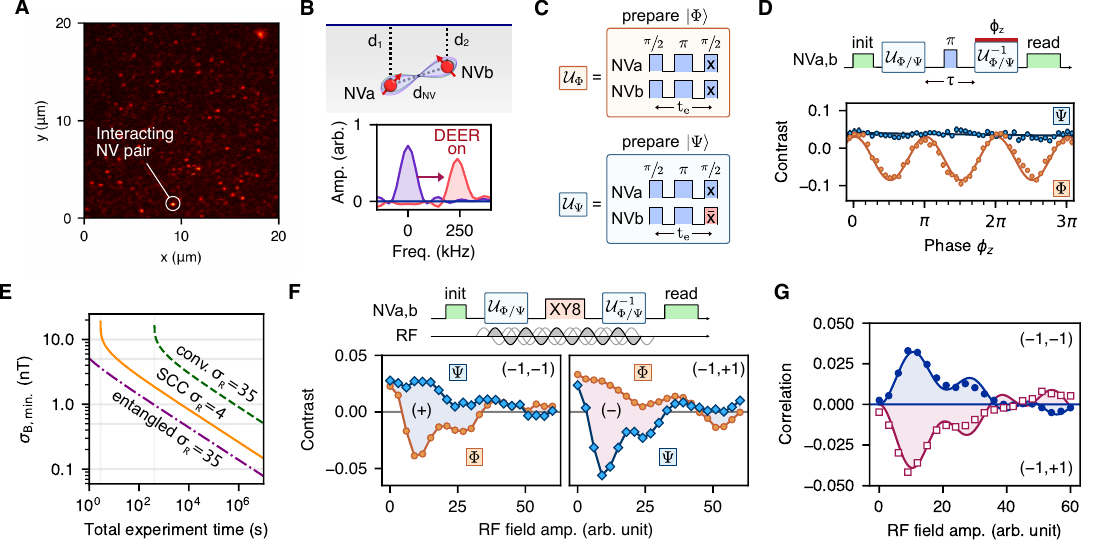}
	\caption{Entanglement-based covariance magnetometry. (A) Representative confocal image of a diamond prepared with N$_2$ molecular ion implantation, with a strongly interacting pair of NV centers indicated. (B) We identify two NV centers spaced $d_\text{NV}\approx 6\,$nm apart and within approximately $10\,$nm of the surface (top), where the dipole-dipole coupling strength is $J_{zz}\approx 2\pi\cdot 250\,$kHz as measured by a DEER experiment on NVa with a recoupling DEER pulse optionally applied to NVb (bottom). (C) Hahn-echo based entanglement blocks to create the $\ket{\Phi}$ (top) and $\ket{\Psi}$ (bottom) Bell states, with pulse spacing $t_\text{e}=\pi/J_{zz}$. (D) (top) Protocol for measuring external fields using the entangled states. (bottom) When we rotate the phases of all pulses in the disentangling block by $\phi_Z$, the $\ket{\Psi}$ state is unaffected (blue) while the $\ket{\Phi}$ state oscillates at $2\phi_Z$ (orange). (E) Calculated sensitivity for measurements of correlated magnetic fields with root-mean-square amplitude $\sigma_B$. Using entangled states with conventional NV center spin readout with readout noise $\sigma_\text{R}=35$ (dot-dashed purple line), we achieve higher sensitivity than is possible with classical states using conventional readout (green dashed line), or with room-temperature SCC readout (see \cite{Supp}). (F) Response of $\ket{\Phi}$ and $\ket{\Psi}$ in the presence of an applied random-phase 1\,MHz external RF field sensed with XY8 decoupling (top). When the same spin transitions are addressed on both NV centers (left), the correlated noise induces a faster response for the $\ket{\Phi}$ state (blue) than the $\ket{\Psi}$ state (orange). When we address opposite spin transitions, the anti-correlated noise induces the opposite effect (right). (G) The difference signal between $S_\Psi$ and $S_\Phi$ reports on the correlation, which is positive when we address the same spin transitions (blue circles) and negative when we address the opposite spin transitions (red squares). Lines are fits to a general model of correlated phase accumulation. In E,F, error bars are smaller than the markers. The NV pair used to acquire the data in F,G is different than the one used for B,D; see \cite{Supp} for details and extended correlation data.
	}
	\label{fig:entanglement_overview}
\end{figure*}

To increase the probability of finding a closely spaced NV center pair while maintaining a sparse distribution of NV centers across the sample area, we fabricate a diamond sample using shallow N$_2$ molecular ion implantation \cite{Gaebel2006,Lee2023} (see \cite{Supp}). In order to identify a pair of entangled NV centers, we use confocal microscope imaging to screen for bright spots that may contain multiple NV centers, and then select candidate pairs that have different crystallographic orientations to allow selective spin control, as well as long enough spin coherence times to enable entanglement. Within a typical $20\,\mu$m $\times$ $20\,\mu$m field of view we observe two or three co-located NV pairs, and occasionally find charge-stable pairs with sufficiently long spin coherence time and sufficiently high dipole-dipole coupling rates to produce entangled states (around 1 in 10, \cref{fig:entanglement_overview}A).  For NV centers with differing orientations, the dipole-dipole interaction takes the form $J_{zz}I_{z_1}I_{z_2}$, as flip-flop terms are suppressed by the detuning of the NV center spin transitions.

In order to perform covariance magnetometry with entangled NV centers, we select a pair of NV centers with coupling strength approximately $J_{zz}=2\pi\cdot250$\,kHz as measured by double electron-electron resonance (DEER) \cite{Neumann2010} (\cref{fig:entanglement_overview}B), corresponding to a spacing of $d_\text{NV}\approx 6\,$nm between the NV centers, comparable to their depth $d\approx 10\,$nm. We then employ a Hahn-echo based entangling gate (\cref{fig:entanglement_overview}C) to create the maximally entangled Bell pairs $\ket{\Phi} = (\ket{0,0} + i \ket{1,1})/\sqrt{2}$ and $\ket{\Psi} = (\ket{0,1} + i \ket{1,0})/\sqrt{2}$, which respond differently to correlated noise: $\ket{\Phi}$ dephases twice as fast as a single-qubit superposition in response to correlated noise, while $\ket{\Psi}$ comprises a decoherence-free subspace in the presence of correlated noise \cite{Bollinger1996,Szankowski2015}. After creating the Bell state, we let it evolve under a conventional dynamical decoupling sequence to detect an external signal, then read out the state by reversing the entanglement gate and subsequently measuring the NV population. This basic measurement protocol is demonstrated in \cref{fig:entanglement_overview}D, which shows the behavior of $\ket{\Phi}$ and $\ket{\Psi}$ in an effective global field, implemented by changing the phase of the disentangling gate. We see that the $\ket{\Phi}$ state accumulates twice this phase, whereas the $\ket{\Psi}$ state is unaffected, as expected. Although the NV-NV coupling is still present during sensing, the dipole coupling term can be effectively removed with a global spin echo after preparing the two NV centers in either Bell state, removing the effect of NV-NV coupling for any AC sensing sequence \cite{Supp}. 

We now utilize this differential response to correlated magnetic fields to perform entanglement-based covariance magnetometry. Defining the correlated signal measured using entangled qubit pairs as $r_\text{e}= (S_\Psi-S_\Phi)/2$, where $S_\Psi$ and $S_\Phi$ are the signals detected using the states $\ket{\Psi}$ and $\ket{\Phi}$ respectively, we find that the correlation can be directly extracted from this difference signal \cite{Supp}:
\begin{align}
    \frac{r_\text{e}}{\sigma_S} &=\sqrt{\frac{2}{1+\sigma_R^2}} \e^{-\chi(t,2t_\text{e})} \braket{\sin[\phiGx(t)]\sin[\phiGy(t)]}, \label{eq:r_entangled}
\end{align}
where $\sigma_S^2$ is the measurement variance from shot noise, the subscripts $a,b$ denote NVa and NVb respectively, the decoherence function $\chi(t,2t_\text{e})$ describes the coherence decay of the NV centers due to the local fields during the sensing time $t$ and both entanglement gate durations $t_\text{e}$ \cite{Cywinski2008}, $\phiGxy$ are the phases accumulated by the NV centers due to the correlated field, and the readout noise $\sigma_{R}=\sqrt{1+2(\alpha_0+\alpha_1)/(\alpha_0-\alpha_1)^2}$ characterizes the fidelity of a photon-counting experiment with mean detected photon number $\alpha_0,\alpha_1$ for spin states $0,1$ respectively \cite{Taylor2008,Hopper2018}. Importantly, \cref{eq:r_entangled} is linear in the readout noise, rather than quadratic as is the case for standard covariance magnetometry using non-interacting NV centers \cite{Rovny2022}, because the presence of a correlated signal is now directly encoded in the final populations of the NV centers. This leads to a signal-to-noise ratio (SNR) gain for entanglement-based covariance magnetometry in the high readout noise regime \cite{Supp}:
\begin{align}
    \frac{\text{SNR}_\text{entangled}}{\text{SNR}_\text{non-interacting}} \approx \sqrt{2}\sigma_R\e^{-\chi_\text{e}(2t_\text{e})}, \label{eq:SNRgain_main}
\end{align}
where $\chi_\text{e}(2t_\text{e})$ is the decoherence function accounting for both the entangling gate and disentangling gate for each NV center. For conventional NV center readout using a green laser at room temperature, $\sigma_R\gtrsim 30$ \cite{Barry2020}, such that the use of entangled states improves the SNR by more than an order of magnitude (\cref{fig:entanglement_overview}E). In practice, this allows us to perform high-sensitivity covariance magnetometry with conventional spin readout using a single green laser, eliminating the need for SCC readout. There are two distinct metrological advantages arising from the use of entanglement. First, in the limit of perfect, quantum-projection-noise-limited readout, $\sigma_R =1$, the $\ket{\Phi}$ state accumulates phase from a correlated signal at twice the rate, which would lead to a $\sqrt{2}$ enhancement in SNR for a field amplitude measurement (Heisenberg scaling). Second, for imperfect readout fidelity, there is an additional advantage, in which the SNR for measuring the degree of correlation improves by a factor of $\sqrt{2}\sigma_R$ by measuring with an intrinsically correlated state. For NV centers in diamond, $\sigma_R$ is typically much higher than the quantum projection noise limit, thus giving rise to a dramatic gain in SNR. We note that for the measurements presented here, we are measuring the degree of correlation in an unknown signal through a difference measurement between Bell pairs, one of which is doubly sensitive to correlated noise and one of which is entirely insensitive. For this difference measurement there is no $\sqrt{2}$ enhancement in the limit of perfect fidelity but there is still a $\sqrt{2}\sigma_R$ enhancement for imperfect sensors \cite{Supp}.

\begin{figure*}[ht]
	\centering
	\includegraphics[width=\textwidth]{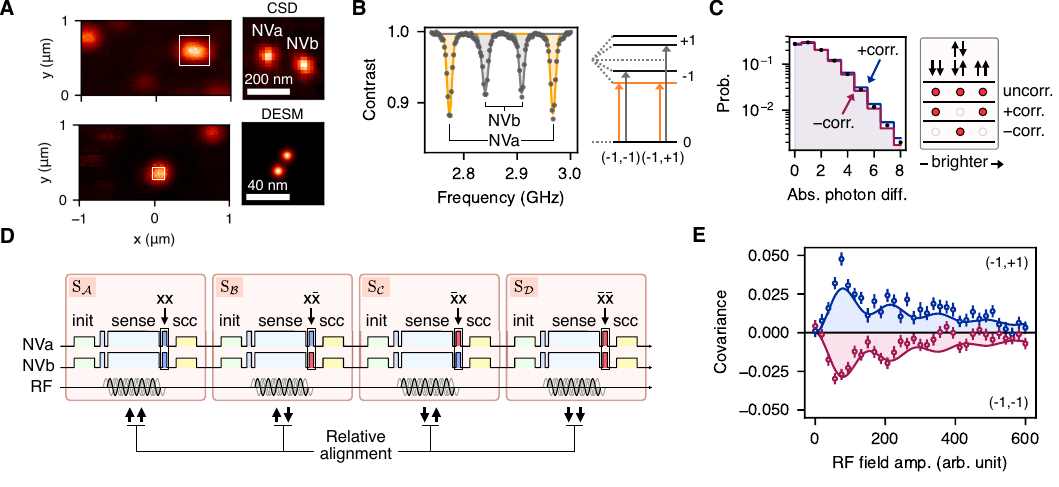}
	\caption{Protocol for detecting correlations below the diffraction limit with resolved spin transitions at low magnetic field. (A) We measure two representative pairs of NV centers with different orientations using confocal scanning (left) and super-resolution imaging (right). Super-resolution imaging is done using purely optical charge state depletion (CSD) when the separation is greater than 30 nm (right, top), or spin-based deterministic emitter-switch microscopy (DESM) when the separation is less than 30 nm (right, bottom). DESM image is reconstructed from Gaussian fitting. (B) Optically detected magnetic resonance spectrum showing two sets of transitions corresponding to NVa (orange) and NVb (gray), with assignments (right). The NV centers are driven independently on either the $m_s=0\rightarrow-1$ transitions for both NV centers, labeled $(-1,-1)$, or the $m_s=0\rightarrow-1$ and $m_s=0\rightarrow+1$ transitions for NVa and NVb respectively, labeled $(-1,+1)$. (C) (left) Probability of absolute photon count difference after phase cycling, $p(|S_A-S_B-S_C+S_D|)$, for NV spin states that are uncorrelated (black circles), positively correlated (blue line), and negatively correlated (red line). (right) The presence of correlations can be determined through the variance of the measured spin-dependent fluorescence. (D) Phase cycling protocol for detecting correlations from two unresolved NV centers. The experiment is repeated four times, where the choice of the relative phase in the final pulse determines the relative spin orientation of the two NV centers, alternately inducing positive and negative correlations. All sequences begin with an initial $y$-phased $\pi/2$ pulse. (E) Covariance measured through phase cycling, under an applied global random-phase RF field with frequency 2\,MHz and varying amplitude. Due to the alignment of the NV centers relative to the external applied field, negative correlations are observed when the same spin transition is address for both NV centers $(-1,-1)$ (red), and positive correlations are observed when we address opposite spin transitions $(-1,+1)$ (blue). Error bars correspond to standard deviation here and throughout all figures unless otherwise stated. Lines are a fit to a phenomenological model that incorporates the expected Bessel function correlation structure \cite{Rovny2022} and an exponential decay. 
	}
	\label{fig:overview}
\end{figure*}

\begin{figure*}[ht]
	\centering
	\includegraphics{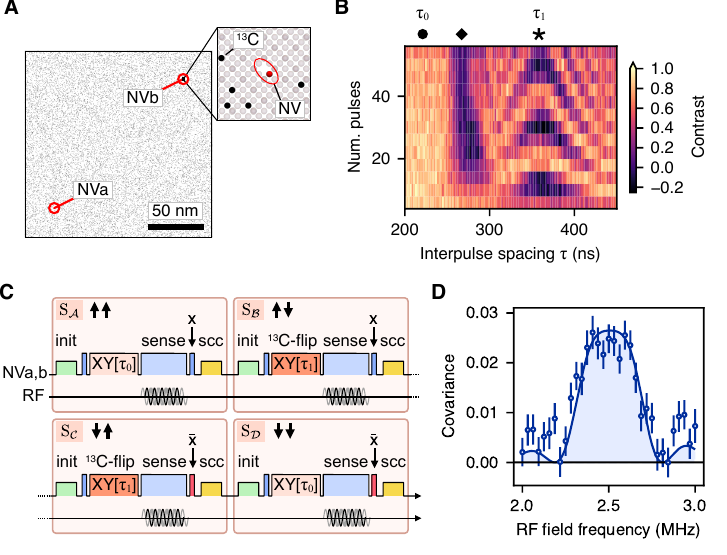}
	\caption{Detecting correlations between optically and spectrally unresolved co-aligned NV centers using a local nuclear spin. (A) Illustration of the typical NV-NV separation and $^{13}$C-NV separation. Shown are a 2D projection of five carbon layers of diamond with two NV centers with the same orientation separated by approximately 80$\,$nm, and randomly populated with a natural $1.1$\% abundance of $^{13}$C nuclei (black). (B) Representative XY-N spectrum of both NV centers using XY dynamical decoupling with N pulses reveals a feature corresponding to the bath of weakly-coupled $^{13}$C (diamond), and a feature corresponding to a nearby $^{13}$C coupled to one NV center (star). As we increase the number of dynamical decoupling pulses resonant with the NV-$^{13}$C coupling, the $^{13}$C-coupled NV center undergoes coherent oscillations due to its coupling with the $^{13}$C spin. (C) Phase cycling protocol using the $^{13}$C to effect selective spin flips. We use an interpulse spacing $\tau_1=358\,$ns when a selective spin flip is required, and an interpulse spacing $\tau_0=220\,$ns otherwise (B, circle). All sequences begin with an initial $y$-phased $\pi/2$ pulse. (D) Covariance detected from an applied random-phase RF tone with a fixed amplitude and swept frequency, using XY16 dynamical decoupling and the $^{13}$C based phase cycling protocol with 10 pulses used to accomplish the selective spin flip. 
	}
	\label{fig:highfield13C}
\end{figure*}

We first demonstrate entanglement-based covariance magnetometry using an external RF coil to apply a global, random phase AC signal as before, which we detect using an XY8 sensing protocol between the two entangling gates. As we increase the amplitude of the RF test signal, $S_\Phi$ responds more quickly than $S_\Psi$, as $\ket{\Phi}$ is more sensitive to correlated noise (\cref{fig:entanglement_overview}F, left). However, when we address opposite spin transitions for each NV center, they experience opposite relative orientations of the RF field and the roles of $\ket{\Phi}$ and $\ket{\Psi}$ are effectively reversed, since $\ket{\Psi}$ will respond more quickly to anti-correlated noise (\cref{fig:entanglement_overview}F, right). The difference signal in each case reports on the correlation $r_\text{e}$, as shown in \cref{fig:entanglement_overview}G. Notably, the data in \cref{fig:entanglement_overview}F and \cref{fig:entanglement_overview}G have markedly higher signal-to-noise ratios compared to \cref{fig:overview}E and \cref{fig:highfield13C}D because of the enhanced sensitivity, while also requiring much less time (about 20 minutes instead of 2 hours for a single data point) and using a single green laser instead of SCC readout. 

\begin{figure*}[ht]
	\centering
	\includegraphics[width=0.97\textwidth]{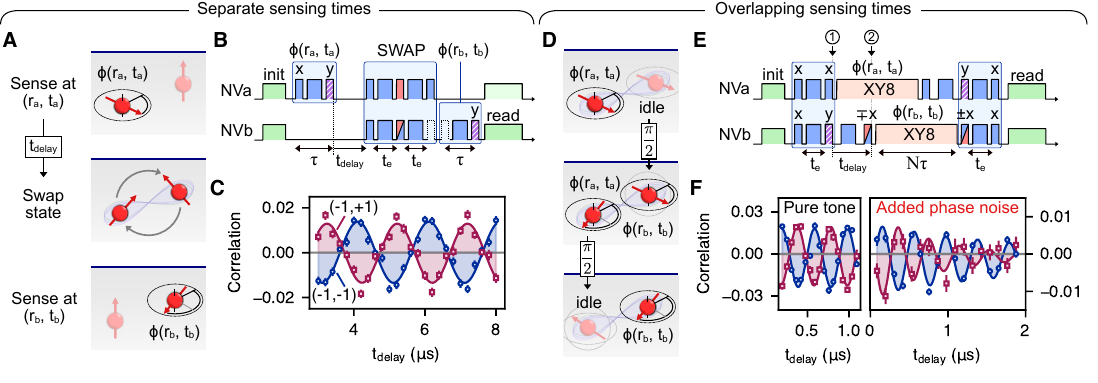}
	\caption{Measuring temporal correlations using strongly coupled NV centers. (A) Measuring correlation between phases accumulated at separate times, adapting a conventional temporal correlation experiment \cite{Laraoui2013} with a SWAP gate transferring the initial phase accumulation from NVa to NVb. (B) Pulse sequence implementation. The SWAP gate incorporates an alternating $\pm x$ phase on the central NVb $\pi/2$ pulse, allowing the correlation to be detected using a subtractive measurement. While the single green laser illuminates both NVa and NVb during readout, only the spin state of NVb is measured, accomplished using a two-cycle contrast measurement. (C) Measurement of coherent time-dependent correlations from an applied 500\,kHz random-phase RF test signal using the sequence in (B), addressing either the same ($(-1,-1)$, blue) or opposite ($(-1,+1)$, red) spin transitions of the two NV centers. (D,E) Protocol for measuring correlation between phases accumulated during partially overlapping sensing times, using Hahn echoes during the delay periods and $N=8$ pulses during sensing. The alternating $\pm x$ phase pulses on NVb allow the correlation to be derived from a subtractive measurement \cite{Supp}. The system is initialized in a superposition of Bell states, and has the idling phase converted to populations after $t_\text{delay}$. (F) Measurement of coherent two-time correlations with a short time delay from a random-phase 2.5\,MHz pure tone (left), and from a random-phase 2.5\,MHz tone with 1\,MHz added phase noise (right), addressing either the same (blue) or opposite (red) spin transitions of the two NV centers. Lines are sinusoidal fits, with frequencies in good agreement with those of the applied random-phase test tones.
	}
	\label{fig:entanglement_timedelay}
\end{figure*}

The interacting, multi-qubit sensor also enables high-sensitivity measurement of nonlocal magnetic fields using gate-based detection sequences that tunably access correlations across space and time. To illustrate this, we now describe two sensing sequences for measuring spatiotemporal correlations using strongly coupled NV centers. The key constraint is to accumulate phase on each NV center at the desired times, while simultaneously removing the effect of the NV-NV coupling during the sensing sequence. The first protocol is applicable when the phase accumulation times of the two NV centers are completely separate; in this case, conventional temporal correlation spectroscopy \cite{Laraoui2013} may be extended to two NV centers by using a SWAP gate during the free evolution time between the first and second phase accumulation blocks (\cref{fig:entanglement_timedelay}A,B). This is sufficient because one or the other NV center will be in the state $m_s=0$ at all sensing times, decoupling the NV centers and removing the need to use entangled states during the sensing itself; however, the use of the SWAP gate still allows the phases to be combined during sensing and before readout, and retains the improved sensitivity from a single readout. While the phases are accumulated using both NVa and NVb, the final signal is encoded entirely in the spin state of NVb, which we selectively measure using selective spin control \cite{Supp}. Finally, we use SWAP gates that alternate the relative spin orientation of NVa and NVb (alternately adding or subtracting their phases), and subtraction reveals the coherent two-time correlation between NVa and NVb produced by a random-phase $500\,$kHz AC signal (\cref{fig:entanglement_timedelay}C). This correlation reverses sign as expected when we repeat the experiment while addressing opposite spin orientations of the two NV centers. 

The second protocol addresses the more challenging case of measuring short-time correlators that require two overlapping phase accumulation periods (\cref{fig:entanglement_timedelay}D,E). We seek to accumulate phase on NVa, then accumulate phase on both NV centers simultaneously, then accumulate phase on NVb (\cref{fig:entanglement_timedelay}D); this is challenging because the sensing pulses on NVa and NVb occur at different times, and no longer remove the NV-NV coupling. 
To accomplish this, we alter the entangling gate to instead initialize the (maximally entangled) Bell pair superposition $\ket{\psi_1}=\tfrac{1}{2}(\ket{0,0}+\ket{1,1}+i\ket{0,1}+i\ket{1,0})$ (\cref{fig:entanglement_timedelay}E, event 1). Then, after the initial phase accumulation time, a $\pi/2$ pulse on NVb restores the $\ket{\Phi}$ (or $\ket{\Psi}$) state for two-qubit sensing, while mapping the unwanted accumulated NVb delay phase into a population difference (\cref{fig:entanglement_timedelay}E, event 2). 
This is repeated at the end of the sensing sequence for the unwanted NVa delay phase, allowing the two-time correlator to be measured at the expense of an added decoherence term from the delay times (see \cite{Supp}): $\tfrac{1}{2}\left(S_\Psi-S_\Phi\right) = \tfrac{1}{2}(\e^{-\chi_{a,\text{delay}}(t)} + \e^{-\chi_{b,\text{delay}}(t)})r_\text{e}$, where $\chi_{a/b,\text{delay}}$ are the decoherence functions incurred by the delay times. This protocol allows us to detect two-time correlators at short separation times (\cref{fig:entanglement_timedelay}F, left), including coherent two-time correlators from signals with very short correlation times (\cref{fig:entanglement_timedelay}F, right).

\section{Conclusions and outlook}

We have demonstrated that multi-qubit sensors, including optically unresolved pairs and clusters with coherent coupling, enable new sensing modalities. In particular, we have shown that intrinsically correlated, maximally entangled states of NV center pairs are ideally suited for measuring magnetic field correlations, giving a large sensitivity boost over measurements with uncorrelated, non-interacting NV centers and offering a useful metrological advantage. These ideas could be extended to larger multi-qubit sensors using circuits designed to create many-body states in dipolar-interacting spin systems \cite{Zheng2022}, which are typically studied for their metrological gain but could also be used to efficiently measure higher-order noise cumulants.

The NV center multi-qubit nanoscale sensor protocols we demonstrated here give access to new physical quantities, especially relevant to a wide range of nonlocal phenomena in many body physics \cite{Rovny2024a}; these include correlated noise statistics from vortex dynamics in superconducting thin films \cite{Curtis2024}, spin fluctuations in magnetic insulators \cite{Chatterjee2019}, nonequilibrium current noise in 2D materials \cite{Zhang2024}, and other quantum phenomena where the characteristic length scales of dynamics are below one micron, including intrinsic noise from the diamond surface itself \cite{Myers2014,Rosskopf2014,Romach2015,Sangtawesin2019}. Looking forward, multiplexed readout of many multi-qubit clusters simultaneously \cite{Cheng2024,Cambria2024} can dramatically improve the throughput of such measurements, while the detection frequencies may be extended from the MHz regime shown here to the GHz regime using correlated $T_1$ measurements. 

Furthermore, these multi-qubit sensing protocols enable new modalities in sensor networks \cite{Brady2024}, in the detection of non-Gaussian statistics \cite{Norris2016}, and in nanoscale measurements of correlations among different components of noise, including among different orientations and different frequencies. More broadly, these concepts are applicable to metrology with entangled sensors and multi-qubit registers in other platforms \cite{Ye2024}, including in trapped atoms for metrology \cite{Schine2022}, in superconducting qubits for measurements of correlated noise \cite{Wilen2021} and rare events \cite{McEwen2022} important for realizing large scale quantum processors, and in long distance networks of entangled atomic clocks \cite{Komar2014}.

\begin{acknowledgments}
We gratefully acknowledge helpful conversations with Eugene Demler, Elisabeth Rulke, E.J.\ Daniels, Trisha Madhavan, Bo Dwyer, Zeeshawn Kazi, Artur Lozovoi, Matt Cambria, Sarang Gopalakrishnan, and Jeff Thompson. This work was supported by the Gordon and Betty Moore Foundation (GBMF12237, DOI 10.37807), the National Science Foundation (Grant No. OMA-2326767), and the Intelligence Community Postdoctoral Research Fellowship Program by the Oak Ridge Institute for Science and Education (ORISE) through an interagency agreement between the US Department of Energy and the Office of the Director of National Intelligence (ODNI) (J.R.).
\textbf{Author contributions:} J.R., S.K., and N.P.d.L.\ developed the theoretical framework for super-resolution magnetometry. J.R.\ carried out covariance magnetometry experiments. J.R., S.K., and N.P.d.L.\ conceived the sensing technique, designed experiments, analyzed the data, and wrote the manuscript.
\textbf{Competing interests:} The authors declare no competing interests. 
\\ \\
\end{acknowledgments}

\putbib[MultiQubitSensing]

\end{bibunit}

\pagebreak
\clearpage

\widetext
\begin{center}
\textbf{\large Supplementary Information}
\end{center}
%%%%%%%%%% Merge with supplemental materials %%%%%%%%%%
%%%%%%%%%% Prefix a "S" to all equations, figures, tables and reset the counter %%%%%%%%%%
\setcounter{secnumdepth}{3}
\setcounter{equation}{0}
\setcounter{figure}{0}
\setcounter{table}{0}
\setcounter{page}{1}
\makeatletter
\renewcommand{\theequation}{S\arabic{equation}}
\renewcommand{\thefigure}{S\arabic{figure}}
%\renewcommand{\bibnumfmt}[1]{[S#1]} % comment for removal
%\renewcommand{\citenumfont}[1]{S#1} % comment for removal
%%%%%%%%%% Prefix a "S" to all equations, figures, tables and reset the counter %%%%%%%%%%
\begin{bibunit}[apsrev4-2] % comment for removal
\setcounter{page}{2} % comment for removal

\section{Methods}

\subsection{Diamond samples and experiment setup}

The diamond sample used for the data shown in \cref{fig:overview} was a $^{12}$C-enriched sample implanted with a nitrogen ion energy of 3 keV, resulting in shallow NV centers roughly $10\,$nm from the surface. The diamond used for the data shown in \cref{fig:highfield13C} was a natural abundance $^{13}$C sample implanted with a nitrogen ion energy of 50 keV, resulting in NV centers approximately $70\,$nm from the surface. The diamond used for the data shown in \cref{fig:entanglement_overview} was a $^{12}$C-enriched sample implanted with N$_2$ molecular ions with an implantation energy of 6 keV, resulting in shallow NV centers approximately $10\,$nm from the diamond surface. The diamond used for the data shown in \cref{fig:entanglement_timedelay} was a natural abundance $^{13}$C sample implanted with N$_2$ molecular ions with an implantation energy of 10 keV, resulting in shallow NV centers approximately $20\,$nm from the diamond surface. 

NV center measurements are performed in a home-built confocal microscope setup. The green illumination is provided by a 532 nm optically pumped solid-state laser (Coherent Sapphire LP 532-300), modulated by an acousto-optic modulator (AOM) (Isomet 1205C-1). The orange readout light for the SCC readout was provided by a 594 nm helium-neon laser (REO 39582) optically modulated with an AOMs (Isomet 1205C-1). The ionization light is provided by an internally modulated 638 nm laser (Hubner Cobolt 06-MLD). For charge state depletion (CSD) measurements, a vortex half wave plate (Thorlabs WPV10L-633) is placed in front of the red excitation laser, as well as a quarter wave plate (Thorlabs WPQ05M-633) to ensure equal excitation of different NV center orientations. 

The green and orange lasers are combined by a 3-channel fiber RGB combiner (Thorlabs RGB26HF), joined with the red laser using a shortpass filter in free space (Thorlabs DMSP605), and scanned using an X-Y galvo mirror (Thorlabs GVS012). A 650 nm longpass dichroic mirror (Thorlabs DMLP650) separates the excitation and collection pathways, and the photoluminescence (PL) is measured by a fiber-coupled avalanche photodiode (Excelitas SPCM-AQRH-16-FC). A Nikon Plan Fluor 100x, NA = 1.30, oil immersion objective is used for focusing the excitation lasers and collecting the PL. For SCC readout, the laser powers used (as measured before the objective) were approximately 3 $\mu$W for orange readout, 100 to 130 $\mu$W for green initialization and readout, and 10 to 30 mW for the red ionization. No shelving pulses were used in the SCC readout protocol.

Microwave pulses are generated using a Rohde and Schwarz signal generator (SMATE200A) and amplified with a high power amplifier (Mini-Circuits ZHL-16W-43S+) before being sent to a homemade microwave stripline. Low frequency test signals are generated with an arbitrary waveform generator (Keysight 33622A), amplified with a high power amplifier (Mini-Circuits LZY-22+), and sent through a homemade RF antenna. All experiments were performed at low external magnetic field, below 100 G, except for the experiments used to collect the data shown in \cref{fig:highfield13C}, which were performed at 1769 G. 

For the data shown in \cref{fig:overview} we apply a $f_0=2\,$MHz AC signal detected using a Hahn echo with sensing time $\tau=250\,$ns. For the data shown in \cref{fig:highfield13C} we use an XY16 sensing sequence to detect a frequency-swept AC RF tone, using pulse timings $\tau=200\,$ns, $\tau_0=220\,$ns, and $\tau_1=358\,$ns, and 10 pulses to accomplish the selective $^{13}$C-mediated spin flips. For the data shown in \cref{fig:entanglement_overview}D, we used time proportional phase incrementation (TPPI) \cite{Marion1983}, incrementing the phase of all pulses in the disentangling block by $\phi_z=\omega_\text{tppi}\tau$, where we chose $\omega_\text{tppi}=2\pi\cdot10\,$MHz. For the entanglement data shown in \cref{fig:entanglement_overview}F,G, we use an entangling gate time of $t_\text{e}=2.732\,\mu$s, and apply a $f_0=1\,$MHz AC RF tone, detected using XY8 with pulse spacing $\tau=500\,$ns. For the data shown in \cref{fig:entanglement_timedelay}C, we use a SWAP gate comprising entangling times $t_e=358\,$ns, and a sensing time $\tau=1\,\mu$s to detect a $f_0=500\,$kHz AC RF tone. For the data shown in \cref{fig:entanglement_timedelay}E we use an entangling gate time $t_\text{e}=2.732\,\mu$s, and an interpulse spacing $\tau=196\,$ns to detect a $f_0=2.5\,$MHz AC RF tone. For the data shown in \cref{fig:entanglement_timedelay}F we use the same parameters, but phase modulate the RF tone using 1\,MHz bandwidth phase noise. Note that different NV pairs are used to acquire the data in \cref{fig:entanglement_overview}, \cref{fig:entanglement_timedelay}C, and \cref{fig:entanglement_timedelay}F.

\subsection{Super-resolution imaging}

We employ both charge-state depletion (CSD) \cite{Chen2015} microscopy for purely optical super-resolution imaging, and deterministic emitter switch microscopy (DESM) for super-resolution imaging based on the NV center spin-dependent fluorescence contrast \cite{Chen2013,Dolde2013}. To perform CSD imaging using the green, red, and orange lasers, with a vortex wave plate in front of the red laser. While we slightly vary the parameters in each experiment, in a typical application we use 5\,$\mu$s of green initializing laser (approximately 150\,$\mu$W), 20\,$\mu$s of red ionizing laser (approximately 30 mW), and 15\,$\mu$s of orange readout laser (approximately 35\,$\mu$W). We repeat this sequence about 10000 times at each pixel. Our resolution is comparable to the value of 28.6 nm reported in \cite{Chen2015} for this ``recharging-CSD'' (rCSD) approach. In the future, ``ionization-CSD'' (iCSD) using a green vortex beam may be implemented to improve this resolution to the reported value of 4.1 nm \cite{Chen2015}. 

For DESM microscopy, we first identify a pair of NV centers with different crystallographic alignments and characterize their Zeeman transitions and $\pi$ pulse times. Then, at each pixel, we perform three interleaved experiments in turn: a $\pi$ pulse on NVa, a $\pi$ pulse on NVb, and a reference experiment with no pulses. The $\pi$ pulses turn each NV center dark: the NVa-only image is therefore constructed from the difference between the third (reference) experiment and the first ($\pi$ on NVa) experiment, and likewise for the NVb-only image \cite{Chen2013,Dolde2013}. The NVa-only and NVb-only images are each fit to a two-dimensional Gaussian to extract the fitted center location. We repeat this experiment approximately 10 times to find the average distance vector, and find the uncertainty from the standard deviation of the distance vector across experiments. Our resolution is approximately 2\,nm in our current implementation, limited by sample drift. 

\section{Variance based super-resolution covariance magnetometry}

\subsection{Basic theory and sensitivity}

Here we assume that each NV center is equally likely to be in state $m_s=0$ or $m_s=1$, and use the generalized readout noise (that is, not assuming Poisson distributed noise) \cite{Rovny2022,Hopper2018}:
\begin{align}
    \sigma^\text{gen}_{R} = 
    \sqrt{1+2\frac{\sigma_0^2 + \sigma_1^2}{(\alpha_0 - \alpha_1)^2}},\label{eq:sigmaRgeneral}
\end{align}
where $\alpha_i$ and $\sigma_i^2$ are the photon mean and variance for the NV center in spin state $m_s=i$. First we write the single-NV photon variance for each NV center individually, using a mixture distribution drawing from spin state 0 with probability $p_0$ and state 1 with probability $p_1$:
\begin{align}
    \sigma^2 = p_0 \sigma_{0}^2 + p_1 \sigma_{1}^2 + p_0 p_1 (\alpha_0-\alpha_1)^2.
\end{align}
For equal probability states $p_0=p_1=0.5$, the single-NV photon standard deviation is 
\begin{align}
    \sigma = \frac{1}{2}(\alpha_0-\alpha_1)\sigma_R^\text{gen}
\end{align}
We now turn to the total variance measured from both NV centers after phase cycling, assuming equal single-NV variances for each cycle for simplicity:
\begin{align}
    \text{Cov} &= \frac{1}{8}\left(\mathcal{A} - \mathcal{B} - \mathcal{C} + \mathcal{D}\right) = \frac{1}{4}(\alpha_0-\alpha_1)^2 r_\text{ideal}. \label{eq:cov_maximum}
\end{align}
Using the gaussian approximation of our photon counts for simplicity, the uncertainty of the measured photon variance is $\mathcal{A}/\sqrt{2N}$, such that the signal to noise ratio is 
\begin{align}
    \text{SNR} \approx \frac{\sqrt{2N}}{r_\text{ideal}+\sigma_R^2}r_\text{ideal} \approx \frac{\sqrt{2N}}{\sigma_R^2}r_\text{ideal}
\end{align}
where $N$ is the number of experiments completed for each phase cycle, and the final approximation holds when $\sigma_R^2\gg r_\text{ideal}$, which is the usual case. Thus the sensitivity is worse than that for covariance magnetometry of resolvable NV centers by a factor of $\sqrt{2}$.

\subsection{Phase cycling and mean subtraction}

The primary challenge of super-resolution covariance magnetometry using variance contrast is that the brightness variance can change for many other reasons, including mean spin value offsets caused by pulse errors or overall brightness changes due to experimental drift (since the mean and variance of the brightness are related). While the phase cycling protocol will remove most of these effects by co-adding the variances in the way described above, we can remove residual baseline offsets by making use of the mean photon counts rather than just the variances. In the case of perfect phase cycling, the net mean photon counts should become zero; when they do not, they report on baseline offsets. 

We use these mean value deviations to subtract baseline offsets by making two simplifying assumptions: (1) that the two NV centers have approximately equal brightnesses, and (2) that the two NV centers have equal mean spin value probabilities. These assumptions are better for the case of two co-aligned NV centers (which will for instance have very similar spin dynamics since pulse errors will be equivalent between the two). Using these assumptions, we can find the baseline expected variances from the measured means since the photon probability distributions are known. First, the mean photon counts and associated spin state probabilities for NV $a$ or $b$ are:
\begin{align}
    \text{Var}(S) &= \sigma_a^2 + \sigma_b^2 + 2r \sigma_a\sigma_b \\
    \mu_{a,b} &= p_0 \alpha_0 + p_1 \alpha_1 \\
    p_0 = \frac{\mu_{a,b}-\alpha_1}{\alpha_0-\alpha_1}& \qquad p_1 = 1-p_0 = \frac{\alpha_0-\mu_{a,b}}{\alpha_0-\alpha_1},
\end{align}
For a single NV center we can write the baseline expected variance in terms of the mean photon counts from the spin measurement, and the calibrated means and variances of the $m_s=0$ and $m_s=\pm1$ spin states:
\begin{align}
    \sigma_{a,b}^2(\mu_{a,b}) = &\left( \frac{\mu_{a,b}-\alpha_1}{\alpha_0-\alpha_1} \right) \sigma_0^2 + \left( \frac{\alpha_0-\mu_{a,b}}{\alpha_0-\alpha_1} \right) \sigma_1^2 + (\mu_{a,b}-\alpha_1)(\alpha_0-\mu_{a,b}) \label{eq:var_from_mean_single}
\end{align}
The key idea of this equation is that the right hand side contains only mean measured photon counts ($\mu_{a,b}$) or calibrated state brightness and variance ($\alpha_i, \sigma^2_i$), which thus allow a measure of the expected baseline photon variance from an experiment ($\sigma_{a,b}^2$) just from the mean values. 

\Cref{eq:var_from_mean_single} describes a calibrated expected photon variance from the measured photon mean for a single NV center. Given the measured signal $S$, which is a sum of the photons from both NV centers and has mean $\mu_S$, we use $\mu_a=\mu_b=\mu_S/2$ and the independently calibrated NV brightness statistics $\alpha_i, \sigma^2_i$ in \cref{eq:var_from_mean_single} to find the expected residual baseline variance $\sigma_a^2=\sigma_b^2\equiv\sigma_{a,b}^2$. We do this for the mean values measured in each phase cycle $\mathcal{A},\mathcal{B},\mathcal{C},\mathcal{D}$, to calculate the phase-cycled baseline variance: 
\begin{align}
    \sigma^2_\text{baseline} = [\sigma_{a,b}^2(\mu_S^\mathcal{A}) - \sigma_{a,b}^2(\mu_S^\mathcal{B}) - \sigma_{a,b}^2(\mu_S^\mathcal{C}) + \sigma_{a,b}^2(\mu_S^\mathcal{D})]/4. 
\end{align}

\section{NV-13C coupling characterization}

In the secular frame of the strong axial zero-field splitting of the NV center in the presence of an aligned external magnetic field, the NV-$^{13}$C Hamiltonian can be written as:
\begin{align}
    \hat{H} = (A_\parallel \hat{I}_z + A_\perp \hat{I}_x)\hat{S}_z + \omega_L \hat{I}_z,
\end{align}
where $\hat{S}$ and $\hat{I}$ are the NV and $^{13}$C spin operators respectively, $A_\parallel$ and $A_\perp$ are the parallel and perpendicular components of the hyperfine coupling, and $\omega_L$ is the $^{13}$C nuclear Larmor frequency. While there are no flip-flop terms in this Hamiltonian, a $^{13}$C-mediated spin flip of the NV center may be accomplished when the NV center is initialized in the transverse plane and then flipped synchronously with the $^{13}$C precession, which we now describe.

\begin{figure*}[ht]
	\centering
	\includegraphics{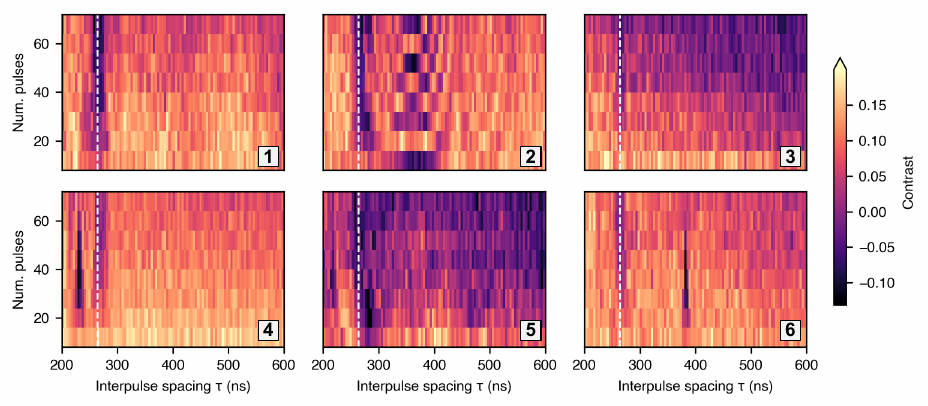}
	\caption{Representative XY spectra for 6 different pairs of co-aligned NV centers, taken at high magnetic field (1769 G). NV center pair 2 was used to acquire the data shown in \cref{fig:highfield13C}. Vertical dashed lines indicate the $^{13}$C bath. Other resonances, including those very close to the bath resonance, indicate strongly coupled $^{13}$C found in this range of XY pulse rate.
	}
	\label{fig:SI_13C_XY8}
\end{figure*}

As shown in \cite{Kolkowitz2012,Taminiau2012,Zhao2012}, the NV center fluorescence signal during an XY-type dynamical decoupling sequence, normalized between $-1$ and $1$ such that the probability of preserving the initial state is $p=(S+1)/2$, is (adapting the notation of \cite{Taminiau2012})
\begin{align}
    S &= 1-\mathcal{N}(\tau) \sin^2\left(\frac{N \phi}{2}\right), \\
    \cos(\phi) &= \cos(\alpha)\cos(\beta) - a_z \sin(\alpha)\sin(\beta), \\
    \mathcal{N}(\tau) &= a_x^2
    \frac{(1-\cos(\alpha))(1-\cos(\beta))}{1+\cos(\alpha)\cos(\beta)-a_z\sin(\alpha)\sin(\beta)},
\end{align}
where we have used
\begin{align}
    \tilde{\omega}^2&\equiv (A_\parallel+\omega_L)^2 + A_\perp^2 \\
    a_x &\equiv \frac{A_\perp}{\tilde{\omega}} \qquad 
    a_z \equiv \frac{A_\parallel + \omega_L}{\tilde{\omega}} \\
    \alpha &\equiv \tilde{\omega}\tau/2 \qquad 
    \beta \equiv \omega_L\tau/2.
\end{align}
In the strong magnetic field limit such that $\omega_L\gg A_\parallel, A_\perp$, the condition for the first resonance such that $\mathcal{N}(\tau_0)=1$ is $\tau_0 = \pi/(4\omega_L + 2A_\parallel)$ (note that our definition of $\tau$ as the interpulse spacing differs from that of \cite{Taminiau2012}). When this condition is satisfied, the NV center will undergo maximum-contrast coherent oscillations as a function of increasing pulse number $N$. Importantly, while the direction of this NV center rotation is conditioned on the orientation of the $^{13}$C, a rotation will occur regardless of the $^{13}$C orientation such that a coherent spin flip becomes possible conditioned only on the presence of a coupled $^{13}$C. 

\begin{figure*}[ht]
	\centering
	\includegraphics[width=0.95\textwidth]{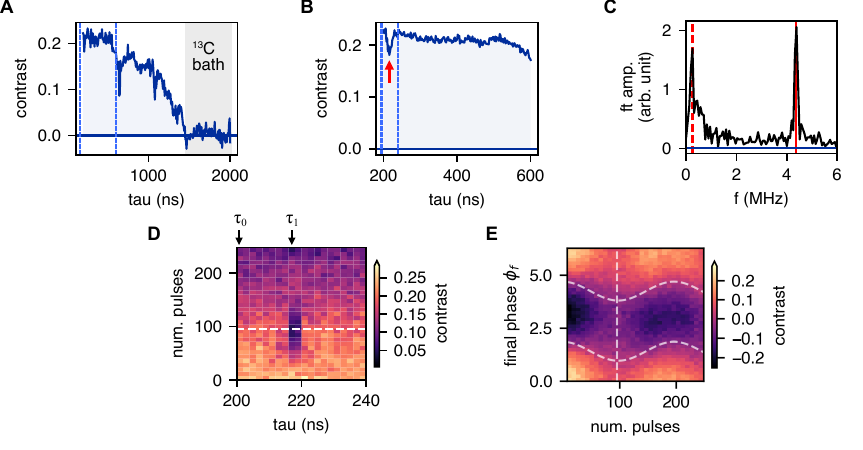}
	\caption{Representative spectroscopy of NV-$^{13}$C coupling. (A) Spectrum using XY8-4 dynamical decoupling, showing the overall decoherence due to the $^{13}$C spin bath. (B) Spectrum using XY8-4 decoupling focusing on the region between the dashed lines in (A). (C) Correlation spectrum derived from the feature indicated by the red arrow in (B). The peak at the $^{13}$C Larmor frequency (red dashed line) identifies this feature as arising from a nearby $^{13}$C nuclear spin, while the other peak (red solid line) indicates the effective frequency enhanced by the NV-$^{13}$C coupling. (D) Pulse-number dependent spectrum of the region between dashed lines in (B), showing coherent NV center oscillations induced by the $^{13}$C used for selective control. The pulse spacing $\tau_0$ is used for the no-spin-flip blocks and $\tau_1$ for the spin flip blocks in the covariance protocol (\cref{fig:highfield13C}D). A spin flip is effected with 96 pulses. (E) Phase deviations during dynamical decoupling. We modify the phase $\phi_f$ of the final $\pi/2$ pulse of an XY sensing sequence, and find a pulse-number dependent phase deviation. This phase deviation is mapped to a final mean-value deviation in the covariance protocol, and is a source of baseline variance fluctuations that must be subtracted as described in the text.
	}
	\label{fig:SI_13C}
\end{figure*}

To run a super-resolution covariance protocol for co-aligned NV centers, we align the external magnetic field, perform an XY8 dynamical decoupling sweep to find resonant responses (\cref{fig:SI_13C}A,B), perform correlation spectroscopy \cite{Laraoui2013} to identify the feature as arising from a $^{13}$C nucleus (\cref{fig:SI_13C}C), then calibrate the $^{13}$C-mediated spin flip by sweeping the interpulse spacing and pulse number (\cref{fig:SI_13C}D). This selective $^{13}$C-mediated spin flip only affects one of the two NV centers, allowing us to prepare different relative NV center alignments and accomplish the phase cycling scheme described above. As an example source of baseline variance fluctuations that must be compensated as described above, we show overall phase oscillations observed as we increase the number of XY pulses (\cref{fig:SI_13C}E).

\section{Entanglement-based covariance magnetometry}

\subsection{The entanglement operation}

In this section we sketch the basic protocol for clarity (e.g.\ focusing on just the $m_s=0$ and $m_s=1$ spin sublevels), and include the full details for the spin-1 system in the next section for completeness. The time evolution operator for the Hahn-echo entanglement gates are
\begin{align}
\mathcal{U}_\Phi &= \mathcal{R}_{\tfrac{\pi}{2}}\,\mathcal{U}_{zz}(t_\text{e}/2)\,\mathcal{R}_{\pi}\,\mathcal{U}_{zz}(t_\text{e}/2)\,\mathcal{R}_{\tfrac{\pi}{2}} \\
\mathcal{U}_\Psi &= \overline{\mathcal{R}}_{\tfrac{\pi}{2}}\,\mathcal{U}_{zz}(t_\text{e}/2)\,\mathcal{R}_{\pi}\,\mathcal{U}_{zz}(t_\text{e}/2)\,\mathcal{R}_{\tfrac{\pi}{2}},
\end{align}
where
\begin{align}
    \mathcal{R}_{\phi} &= \text{exp}\left[-i (I_{x_1} + I_{x_2}) \phi \right] \\
    \overline{\mathcal{R}}_{\phi} &= \text{exp}\left[-i (I_{x_1} - I_{x_2}) \phi \right] \\
    \mathcal{U}_{zz}(t) &= \text{exp}\left[-i J_{zz}\, I_{z_1}I_{z_2} t \right] \\
    t_\text{e} &= \frac{\pi}{J_{zz}},
\end{align}
$J_{zz}$ is the NV-NV coupling strength, and $t_\text{e}$ is the time to create a maximally entangled state. Writing the entanglement propagator for just the $m_s=0$ and $m_s=1$ spin sublevels, we find
\begin{align}
\mathcal{U}_\Phi &\: \dot{=} \: \frac{\e^{-i\pi/4}}{\sqrt{2}}
\begin{pNiceMatrix}[first-row, code-for-first-row = \color{blue}, cell-space-limits = 2pt,
rules/color=[gray]{0.9},rules/width=1pt]
\ket{1,1} & \ket{1,0} & \ket{0,1} & \ket{0,0} \vspace{1mm} \\ \hline
1 & 0 & 0 & +i \\
0 & 1 & -i & 0 \\
0 & -i & 1 & 0 \\
+i & 0 & 0 & 1
\end{pNiceMatrix} \\[10pt]
\mathcal{U}_\Psi &\: \dot{=} \: \frac{\e^{-i\pi/4}}{\sqrt{2}}
\begin{pNiceMatrix}[first-row, code-for-first-row = \color{blue}, cell-space-limits = 2pt,
rules/color=[gray]{0.9},rules/width=1pt]
\ket{1,1} & \ket{1,0} & \ket{0,1} & \ket{0,0} \\ \hline
0 & i & 1 & 0 \\
i & 0 & 0 & -1 \\
-1 & 0 & 0 & i \\
0 & 1 & i & 0
\end{pNiceMatrix},
\end{align}
where we have indicated the basis above each matrix representation for clarity. Thus, for an initially polarized spin state $\ket{m_{s_1},m_{s_2}}=\ket{0,0}$, the prepared entangled states are 
\begin{align}
\ket{\Phi(0)} \equiv \mathcal{U}_\Phi \ket{00} = &\frac{\text{e}^{-i \pi/4}}{\sqrt{2}}\left(\ket{0,0} + i \ket{1,1}\right) \\
\ket{\Psi(0)} \equiv \mathcal{U}_\Psi \ket{00} = &\frac{\text{e}^{+i \pi/4}}{\sqrt{2}}\left(\ket{0,1} +  i \ket{1,0}\right).
\end{align}
Importantly, the NV-NV coupling Hamiltonian is removed for any sensing protocol that uses a global $\pi$ spin echo pulse (or multiple pulses), since the $\ket{\Phi}$ state will acquire phase on only the $\ket{1,1}$ vector, while the $\ket{\Psi}$ state is unaffected by the coupling.

\subsection{Sensing fields with the entangled state}

For sensing, we create an entangled state, let it acquire phase in the presence of external fields, then reverse the entanglement gate to map the phases to measurable state populations. After acquiring phase under the external fields to be sensed, the entangled states become 
\begin{align}
\ket{\Phi(t)} = \mathcal{U}_\text{ext}(t)\ket{\Phi(0)} = \frac{\text{e}^{-i \pi/4}}{\sqrt{2}}&\left(\ket{0,0} + i \e^{-i(\phi_a + \phi_b)}\ket{1,1}\right) \\
\ket{\Psi(t)}= \mathcal{U}_\text{ext}(t)\ket{\Psi(0)} = \frac{\text{e}^{+i \pi/4}}{\sqrt{2}}&\left(\ket{0,1} +  i \e^{-i(\phi_a - \phi_b)} \ket{1,0}\right)\text{e}^{-i \phi_b},
\end{align}
where we have pulled out the phase acquired by NV b to make the comparison more clear. 

The dis-entangling operators are similar to the entangling operators:
\begin{align}
\overline{\mathcal{U}}_{\Phi} &= \mathcal{R}_{-\tfrac{\pi}{2}}\,\mathcal{U}_{zz}(t_\text{e}/2)\,\mathcal{R}_{\pi}\,\mathcal{U}_{zz}(t_\text{e}/2)\,\mathcal{R}_{-\tfrac{\pi}{2}}
\: \dot{=} \: \frac{\e^{-i\pi/4}}{\sqrt{2}}
\begin{pNiceMatrix}[first-row, code-for-first-row = \color{blue}, cell-space-limits = 2pt,
rules/color=[gray]{0.9},rules/width=1pt]
\ket{1,1} & \ket{1,0} & \ket{0,1} & \ket{0,0} \\ \hline
1 & 0 & 0 & +i \\
0 & 1 & -i & 0 \\
0 & -i & 1 & 0 \\
+i & 0 & 0 & 1
\end{pNiceMatrix}\\
\overline{\mathcal{U}}_{\Psi} &= \mathcal{R}_{+\tfrac{\pi}{2}}\,\mathcal{U}_{zz}(t_\text{e}/2)\,\mathcal{R}_{\pi}\,\mathcal{U}_{zz}(t_\text{e}/2)\,\overline{\mathcal{R}}_{-\tfrac{\pi}{2}}
\: \dot{=} \: \frac{\e^{-i\pi/4}}{\sqrt{2}}
\begin{pNiceMatrix}[first-row, code-for-first-row = \color{blue}, cell-space-limits = 2pt,
rules/color=[gray]{0.9},rules/width=1pt]
\ket{1,1} & \ket{1,0} & \ket{0,1} & \ket{0,0} \\ \hline
0 & -1 & i & 0 \\
1 & 0 & 0 & i \\
i & 0 & 0 & 1 \\
0 & i & -1 & 0
\end{pNiceMatrix}.
\end{align}
The final states prior to readout are then:
\begin{align}
    \ket{\phi(t)} &= \overline{\mathcal{U}}_\Phi \mathcal{U}_\text{ext}(t) \mathcal{U}_\Phi \ket{00}
    = -i\e^{-i\frac{\phi_a+\phi_b}{2}} \left[\cos\left(\frac{\phi_a+\phi_b}{2}\right)\ket{0,0} - \sin\left(\frac{\phi_a+\phi_b}{2}\right)\ket{1,1}\right] \label{eq:statePhi} \\
    \ket{\psi(t)} &= \overline{\mathcal{U}}_\Psi \mathcal{U}_\text{ext}(t) \mathcal{U}_\Psi \ket{00}
    = -\e^{-i\frac{\phi_a+\phi_b}{2}} \left[\cos\left(\frac{\phi_a-\phi_b}{2}\right)\ket{0,0} + \sin\left(\frac{\phi_a-\phi_b}{2}\right)\ket{1,1}\right]. \label{eq:statePsi}
\end{align}

\subsection{Measuring correlations using the entangled state}

\Cref{eq:statePhi,eq:statePsi} show that the state of the overall system mimics that of a single NV center in a Ramsey-type protocol detecting phase $\phi_a\pm \phi_b$. Namely, for perfect readout of the spin states $\braket{m}=\braket{m_{s_a}+m_{s_b}}$ with probability $p(m_{s_a},m_{s_b})$ we find
\begin{align}
    S_\Phi^\text{ideal} = (1+1)p_\Phi(1,1) &= 2\braket{\sin^2\left(\frac{\phi_a+\phi_b}{2}\right)} = 1 - \braket{\cos(\phi_a)\cos(\phi_b)} + \braket{\sin(\phi_a)\sin(\phi_b)}  \\
    S_\Psi^\text{ideal} = (1+1)p_\Psi(1,1) &= 2\braket{\sin^2\left(\frac{\phi_a-\phi_b}{2}\right)} = 1 - \braket{\cos(\phi_a)\cos(\phi_b)} - \braket{\sin(\phi_a)\sin(\phi_b)}.
\end{align}
Then the difference signal is 
\begin{align}
    \frac{1}{2}\left(S_\Phi^\text{ideal}-S_\Psi^\text{ideal}\right) = \braket{\sin(\phi_a)\sin(\phi_b)},
\end{align}
where the phases $\phi_a,\phi_b$ still represent the total phases acquired by the NV centers (including both correlated and uncorrelated sources). Separating the contributions from correlated ($\phi_{C_i}$) and uncorrelated (or ``local'', $\phi_{L_i}$) sources for NV center $i$ \cite{Rovny2022}, we find
\begin{align}
    \frac{1}{2}\left(S_\Phi^\text{ideal}-S_\Psi^\text{ideal}\right) &= 
    \e^{-\chi_\text{e}(2t_\text{e})}\e^{-[\tilde{\chi}_a(t)+\tilde{\chi}_b(t)]} \braket{\sin[\phiGx(t)]\sin[\phiGy(t)]} \\
    &= \e^{-\chi_\text{e}(2t_\text{e})}r_\text{ideal} \label{eq:rideal} \\
    &\equiv r_\text{e,ideal}
\end{align}
where $\braket{\cos(\phiL)} = \braket{\e^{i\phiL}} = \e^{-\tilde{\chi}(t)}$ is the decoherence function for variance detection \cite{Degen2017}, $\e^{-\chi_\text{e}(2t_\text{e})} = (\e^{-\chi_\text{e,a}(2t_\text{e})}+\e^{-\chi_\text{e,b}(2t_\text{e})})/2$ is the decoherence function for each pair of entangling gates (e.g.\ $\e^{-\chi_\text{e,a}(2t_\text{e})}$ from the entangling and disentangling gates for NVa, where each gate has duration $t_\text{e}$), and $\rideal$ is the NV-NV correlation that would be measured for perfect readout of two NV centers \cite{Rovny2022}. Thus the two-NV correlation is directly accessible from the behavior of $\ket{\Phi}$ and $\ket{\Psi}$, without requiring measurements of products of signals, at the cost of decoherence during the entangling and disentangling gates. We discuss the sensitivity ramifications below.

\subsection{Measuring two-time correlations}

The two-time experiments in \cref{fig:entanglement_timedelay}B,C are formally identical to a conventional correlation spectroscopy experiment \cite{Laraoui2013}, except that the first phase acquisition is completed on the first NV center and the second phase acquisition is completed on the second center. Thus, \cref{eq:rideal} results. Further, since the idling NV center is always in state $m_s=0$, no NV-NV coupling is effective during the sensing steps. The SWAP gate is incorporated as two concatenated CNOT gates, omitting one CNOT operation because NVb begins in the known state $m_s=0$. We also omit a final $\pi/2$ pulse in the SWAP gate, since NVb immediately begins a phase accumulation step.

\begin{figure*}[ht]
	\centering
	\includegraphics[width=0.8\textwidth]{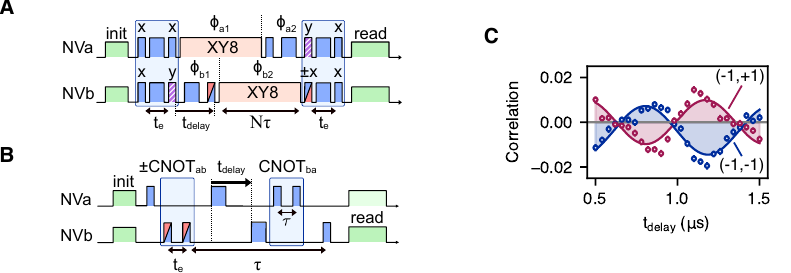}
	\caption{(A) Sequence labeling for measuring two-time correlations. (B) Example of an alternative pulse sequence for measuring two-time correlations using CNOT gates. (C) Results of using the sequence in (B). Systematic artifacts overlay the oscillations. We attribute this to dynamics during the CNOT gate durations ($t_\text{e}=358\,$ns), which occupy a non-negligible fraction of the phase accumulation time ($\tau=4\,\mu$s) and incur hyperfine interactions with the $^{14}$N nuclear spin intrinsic to each NV center.
	}
	\label{fig:SI_OverlapTiming}
    
\end{figure*}

For the two-time experiments in \cref{fig:entanglement_timedelay}D,E, we denote the phases accumulated on NVa,b as shown in \cref{fig:SI_OverlapTiming}A (note that $\phi_{a2}\equiv\phi_{a,\text{delay}}$ and $\phi_{b1}\equiv\phi_{b,\text{delay}}$, defined differently here for notational convenience). After the first delay period, during which NVb collect the unwanted phase $\phi_{b1}$,
a $\pi/2$ pulse on NVb restores the $\ket{\Phi}$ state for two-qubit sensing, while mapping the NVb delay phase $\phi_{b,\text{delay}}$ into a population difference, partially mixing the $\ket{\Phi}$ state into $\ket{\Psi}$ (\cref{fig:entanglement_timedelay}E, event 2):
\begin{align}
    \ket{\psi_2}=&\tfrac{1}{2}\cos\left(\tfrac{\phi_{b,\text{delay}}}{2}\right)(\ket{0,0}+\e^{-i\phi_a}\ket{1,1})  \nonumber\\
    - &\tfrac{1}{2}\sin\left(\tfrac{\phi_{b,\text{delay}}}{2}\right)(\ket{0,1}+\e^{-i\phi_a}\ket{1,0}). 
\end{align}
This takes the appearance of a state that has only acquired phase on NVa, while the delay phase has been converted from a measurable phase into a decoherence mechanism. This is repeated at the end of the sequence for the unwanted NVa delay phase $\phi_{a2}$.
At the end of the sequence, the final states are 
\begin{align}
    \ket{\phi(t)} &= \e^{-\frac{i}{2}(\phi_{a1}+\phi_{a2}+\phi_{b1}+\phi_{b2})} \times \nonumber \\
    \Bigl[+&\left\{ \cosp[\frac{\phi_{a2}+\phi_{b1}}{2}]\sinp[\frac{\phi_{a1}}{2}]\cosp[\frac{\phi_{b2}}{2}] 
    + \cosp[\frac{\phi_{a2}-\phi_{b1}}{2}]\cosp[\frac{\phi_{a1}}{2}]\sinp[\frac{\phi_{b2}}{2}] \right\} \ket{0,0} \nonumber \\
    -&\left\{ \sinp[\frac{\phi_{a2}+\phi_{b1}}{2}]\sinp[\frac{\phi_{a1}}{2}]\cosp[\frac{\phi_{b2}}{2}] 
    + \sinp[\frac{\phi_{a2}-\phi_{b1}}{2}]\cosp[\frac{\phi_{a1}}{2}]\sinp[\frac{\phi_{b2}}{2}] \right\} \ket{0,1}  \nonumber \\
    -&\left\{ \cosp[\frac{\phi_{a1}+\phi_{b2}}{2}]\sinp[\frac{\phi_{a2}}{2}]\cosp[\frac{\phi_{b1}}{2}] 
    - \cosp[\frac{\phi_{a1}-\phi_{b2}}{2}]\cosp[\frac{\phi_{a2}}{2}]\sinp[\frac{\phi_{b1}}{2}] \right\} \ket{1,0} \nonumber \\
    +&\left\{ \cosp[\frac{\phi_{a1}+\phi_{b2}}{2}]\cosp[\frac{\phi_{a2}}{2}]\cosp[\frac{\phi_{b1}}{2}] 
    + \cosp[\frac{\phi_{a1}-\phi_{b2}}{2}]\sinp[\frac{\phi_{a2}}{2}]\sinp[\frac{\phi_{b1}}{2}] \right\} \ket{1,1}\Bigr]  \label{eq:statePhi2T} \\
    \ket{\psi(t)} &= \e^{-\frac{i}{2}(\phi_{a1}+\phi_{a2}+\phi_{b1}+\phi_{b2})} \times \nonumber \\
    \Bigl[&\left\{ \cosp[\frac{\phi_{a2}+\phi_{b1}}{2}]\sinp[\frac{\phi_{a1}}{2}]\cosp[\frac{\phi_{b2}}{2}] 
    - \cosp[\frac{\phi_{a2}-\phi_{b1}}{2}]\cosp[\frac{\phi_{a1}}{2}]\sinp[\frac{\phi_{b2}}{2}] \right\} \ket{0,0} \nonumber \\
    -&\left\{ \sinp[\frac{\phi_{a2}+\phi_{b1}}{2}]\sinp[\frac{\phi_{a1}}{2}]\cosp[\frac{\phi_{b2}}{2}] 
    - \sinp[\frac{\phi_{a2}-\phi_{b1}}{2}]\cosp[\frac{\phi_{a1}}{2}]\sinp[\frac{\phi_{b2}}{2}] \right\} \ket{0,1}  \nonumber \\
    -&\left\{ \cosp[\frac{\phi_{a1}-\phi_{b2}}{2}]\sinp[\frac{\phi_{a2}}{2}]\cosp[\frac{\phi_{b1}}{2}] 
    - \cosp[\frac{\phi_{a1}+\phi_{b2}}{2}]\cosp[\frac{\phi_{a2}}{2}]\sinp[\frac{\phi_{b1}}{2}] \right\} \ket{1,0} \nonumber \\
    +&\left\{ \cosp[\frac{\phi_{a2}-\phi_{b1}}{2}]\cosp[\frac{\phi_{a1}}{2}]\cosp[\frac{\phi_{b2}}{2}] 
    + \cosp[\frac{\phi_{a2}+\phi_{b1}}{2}]\sinp[\frac{\phi_{a1}}{2}]\sinp[\frac{\phi_{b2}}{2}] \right\} \ket{1,1}\Bigr]. \label{eq:statePsi2T}
\end{align}

Taking the difference signal, we find
\begin{align}
    \frac{1}{2}\left(S_\Psi^\text{ideal}-S_\Phi^\text{ideal}\right) = 
    \left\langle\frac{\cosp[\phi_{a2}]+\cosp[\phi_{b1}]}{2}\sinp[\phi_{a1}]\sinp[\phi_{b2}]\right\rangle,
\end{align}
which for small acquired phases during the delay times is
\begin{align}
    \frac{1}{2}\left(S_\Psi^\text{ideal}-S_\Phi^\text{ideal}\right) &\approx
    \frac{1}{2}(\e^{-\chi_{a,\text{delay}}(t)} + \e^{-\chi_{b,\text{delay}}(t)})\e^{-[\tilde{\chi}_a(t)+\tilde{\chi}_b(t)]} \braket{\sin[\phiGx(t)]\sin[\phiGy(t)]}\nonumber\\
    &= \frac{1}{2}(\e^{-\chi_{a,\text{delay}}(t)} + \e^{-\chi_{b,\text{delay}}(t)})\rideal,
\end{align}
where $\chi_{a/b,\text{delay}}$ are the decoherence functions incurred during both the delay periods and entangling gates as described above.

Alternative pulse sequences may be used to access the same information measured by the pulse sequence in \cref{fig:SI_OverlapTiming}A, and different sequences may be optimal in different circumstances. For instance, the sequence shown in \cref{fig:SI_OverlapTiming}B avoids the extra decoherence incurred during the delay periods using CNOT gates mid-circuit; however, because the CNOT gate durations are non-negligible, this sequence can result in systematic deviations which we attribute to e.g.\ hyperfine interactions with the $^{14}$N nuclear spin intrinsic to each NV center during the CNOT operation (\cref{fig:SI_OverlapTiming}C). In general, the pulse sequence used to measure two-time correlators with short delays must simultaneously decouple the NV-NV interaction and measure the phase from the sensing target.

\subsection{Sensitivity of entanglement-based covariance magnetometry}

In this section we will assume the simplest case of two NV centers with equal brightness, with Poisson-distributed photon counts having mean values $\alpha_0$ in spin state $m_s=0$ and $\alpha_1$ in spin state $m_s=1$. We begin by calculating the mean number of detected photons from a measurement of the states described in \cref{eq:statePhi,eq:statePsi}:
\begin{align}
    S_{\Phi/\Psi} = \sum_{\nx,\ny} (\nx + \ny) \big[ 
    &P(\nx|m_s\seq0)P(\ny|m_s\seq0) \braket{\cos^2{\left(\frac{\phi_a\pm\phi_b}{2}\right)} } + \nonumber \\
    &P(\nx|m_s\seq1)P(\ny|m_s\seq1) \braket{\sin^2{\left(\frac{\phi_a\pm\phi_b}{2}\right)} } \big] \\
     = (\alpha_1 - \alpha_0) S_{\Phi/\Psi}^\text{ideal}& + 2\alpha_0.
\end{align}
Then the difference signal is
\begin{align}
    r_\text{e}=\frac{1}{2}\left(S_\Psi - S_\Phi \right) = (\alpha_0-\alpha_1)r_\text{e,ideal}. \label{eq:Sideal}
\end{align}
Note that the sign difference between \cref{eq:Sideal,eq:rideal} is an unimportant convention simply due to the choice of spin states. Finally, to make clear the signal-to-noise ratio and complete the comparison with the detection of two-point correlators from resolvable NV centers \cite{Rovny2022}, we normalize this result by the standard deviation of photons $\sigma_S$. Because we assume equal NV center brightness, and assume measurements of $\ket{0,0}$ or $\ket{1,1}$ (\cref{eq:statePhi,eq:statePsi}), this standard deviation will be identical to that for a single NV center with brightness $2\alpha_0$ for spins states $m_s=0$ and $2\alpha_1$ for spin states $m_s=1$ \cite{Rovny2022}:
\begin{align}
    \sigma_S^2 = \braket{S^2}-\braket{S}^2 = \frac{1}{4}\left(2\alpha_0 - 2\alpha_1 \right)^2 + \frac{1}{2}\left(2\alpha_0 + 2\alpha_1 \right) = \frac{1}{2}(\alpha_0-\alpha_1)^2(1+\sigma_R^2). 
\end{align}
Then the normalized difference signal is
\begin{align}
    \frac{r_\text{e}}{\sigma_S} = \frac{\sqrt{2}}{\sqrt{1+\sigma_R^2}}r_\text{e,ideal}, \label{eq:ridealSigmaR}
\end{align}
and the signal to noise ratio may be compared to that for a covariance measurement detecting two resolvable NV centers (assuming both NV centers have the same decoherence during the entangling gate for simplicity, such that $r_\text{e,ideal}=\e^{-\chi_\text{e}(2t_\text{e})}r_\text{ideal}$) \cite{Rovny2022}:
\begin{align}
    \text{SNR}_\text{entangled} &= \frac{\sqrt{2N}\e^{-\chi_\text{e}(2t_\text{e})}}{\sqrt{1+\sigma_R^2}}r_\text{ideal} \label{eq:SNRentangled} \\
    \text{SNR}_\text{non-interacting} &= \frac{\sqrt{N}}{\sigma_R^2}r_\text{ideal}, \label{eq:SNRclassical}
\end{align}
where $N$ is the total number of experiments (such that there are e.g.\ $N/2$ experiments measuring $S_\Phi$). Note that while \cref{eq:SNRentangled} is strictly defined for $N$ total measurements, the definition of the ratio in \cref{eq:ridealSigmaR} is chosen to have $\sigma_S$ in the denominator (as opposed to the standard deviation of $r_\text{e}$) to make the ratio equivalent to $1/\sqrt{N}$ of the SNR, and to saturate at 1 for a perfect correlation measured with perfect readout $\sigma_R=1$. Crucially, \cref{eq:SNRentangled} is linear in $\sigma_R^{-1}$ rather than quadratic, such that entanglement-based covariance magnetometry is far more sensitive in the low-fidelity limit, which is the usual case for NV centers. In this low-fidelity limit $\sigma_R\gg1$, the signal-to-noise enhancement is 
\begin{align}
    \frac{\text{SNR}_\text{entangled}}{\text{SNR}_\text{non-interacting}} \approx \sqrt{2}\sigma_R\e^{-\chi_\text{e}(2t_\text{e})}.
\end{align}
For conventional readout of NV centers based on spin-dependent fluorescence under green illumination (in a conventional oil microscope without the use of microstructures, which is our current case), the readout noise is $\sigma_R \sim 30$ \cite{Rovny2022,Barry2020}, corresponding to more than an order of magnitude improvement from the entanglement protocol. In the limit of perfect readout fidelity $\sigma_R =1$ (and ignoring decoherence during the entangling gates), $\text{SNR}_\text{entangled}$ becomes identical to $\text{SNR}_\text{non-interacting}$. This is because rather than measuring the amplitude of a correlated signal, we are measuring the degree of correlation in an unknown signal. A measurement designed to determine the amplitude of a correlated signal would not require the $N/2$ experiments prepared in the $\ket{\Psi}$ state, and would thus recover the two-qubit Heisenberg sensitivity scaling.

In practice, we repeat each experiment twice with difference choices for the phase of the final pulse, resulting in signal and reference data
\begin{align}
    S_{\Phi/\Psi}^\text{sig} &= (\alpha_1 - \alpha_0) S_{\Phi/\Psi}^\text{ideal} + 2\alpha_0 \\
    S_{\Phi/\Psi}^\text{ref} &= (\alpha_0 - \alpha_1) S_{\Phi/\Psi}^\text{ideal} + 2\alpha_1.
\end{align}
We then calculate the signal contrast defined by
\begin{align}
    S_{\Phi/\Psi}^\text{cont.} = \frac{S_{\Phi/\Psi}^\text{sig}-S_{\Phi/\Psi}^\text{ref}}{S_{\Phi/\Psi}^\text{sig} + S_{\Phi/\Psi}^\text{ref}}.
\end{align}
It is straightforward to show that the difference signal is then
\begin{align}
    \frac{1}{2}\left( S_{\Psi}^\text{cont.} - S_{\Phi}^\text{cont.} \right) = \left( \frac{\alpha_0-\alpha_1}{\alpha_0+\alpha_1} \right) r_\text{e,ideal}
\end{align}
which may be used to extract the correlation from the independently calibrated NV center brightnesses $\alpha_0$ and $\alpha_1$.

\subsection{Sensitivity of entanglement-based covariance spectroscopy}

The coherence decay and spectral density are described as follows:
\begin{align}
     C(t) &= \e^{-\chi(t)}, \qquad \chi(t) = \frac{1}{2}\braket{\phi^2} = 
     \frac{1}{\pi}\int_0^{\infty} d\omega \frac{F(\omega)}{\omega^2}S(\omega)
     \approx \frac{t}{\pi} S(\omega), \qquad S(\omega) = \int_{-\infty}^{\infty}\e^{-i\omega t}\gamma_e^2 \braket{B(t'+t)B(t')} dt \label{eq:spectralbasics}
\end{align}
where $F(\omega)$ is the pulse sequence filter function, $\chi(t)$ is the decoherence from all noise sources (correlated and uncorrelated), and $S(\omega)$ is the spectral density of the magnetic field \cite{Degen2017,Szankowski2017,Romach2015}.

To derive the correlated spectrum \cite{Rovny2022}, we make three simplifying assumptions: (1) that the noise spectrum may be decomposed into correlated and uncorrelated (local) contributions $S(\omega) = S_C(\omega)+S_{L_{1,2}}(\omega)$, (2) that the correlated component of the noise is identical at the locations of the two NV centers with noise spectral densities $S_C(\omega) = S_{C_1}(\omega) = S_{C_2}(\omega)$, and (3) that the accumulated correlated phases are Gaussian-distributed or small such that $\phi\ll\pi$. Then we have $\braket{\sin(\phiGx)\sin(\phiGy)}=\e^{-2\chi_C}\sinh(2\chi_C)$, where $\chi_C$ is the decoherence induced by the correlated noise source, and the correlated spectral density becomes
\begin{align}
    S_C(\omega) = \frac{\pi}{2t}
    \sinh^{-1}\left(\frac{\rideal}{C_1(t) C_2(t)}\right)  \label{eq:reconstructedSpectrumSupp}
\end{align}
where $C(t)=\e^{-(\tilde{\chi}+\chi_C)t}$ is the total decoherence from all sources and $\rideal$ is given by \cref{eq:ridealSigmaR}.

Using the SNR derived above (\cref{eq:SNRentangled}), we may now extend the results from \cite{Rovny2022} to find the sensitivity straightforwardly. Assuming the noise has flat spectral density around the detection frequency $S_{C}(\omega)=\gamma_e^2\sigma_B^2/\text{Hz}$ we find the minimum detectable noise amplitude
\begin{align}
    \sigma_{B,\text{min}}^2 &= \frac{-\pi\cdot\text{Hz}}{4\gamma_e^2t} \ln \left( 1-\frac{\sqrt{2} \sigma_R \e^{2(t+2t_\text{e})/T_2}}{\sqrt{T/(t+2t_\text{e}+t_\text{R})}} \right) \approx 
    \frac{\pi\cdot\text{Hz}}{\gamma_e^2t} \frac{\sigma_R \e^{2(t+2t_\text{e})/T_2}}{\sqrt{2T/(t+2t_\text{e}+t_\text{R})}}, \label{eq:sensitivitygeneralSupp}
\end{align}
where $t$ is the phase integration time, $t_\text{R}$ is the readout time, $t_\text{e}$ is the entanglement gate duration, $T\approx (t+2t_\text{e}+t_\text{R})N$ is the total experiment time ignoring initialization, and the approximation holds for large experiment numbers $N \gg 1$. \Cref{eq:sensitivitygeneralSupp} is shown in \cref{fig:entanglement_overview}E assuming a coherence time $T_2=100\,$ $\mu$s, an entangling gate time $t_\text{e}=2\,\mu$s, a phase integration time $t=T_2/4=25\,\mu$s, and a readout time of $300\,$ns for conventional readout and $1\,$ms for SCC readout. Initialization time (typically a few microseconds) was ignored.

\begin{figure*}[ht]
	\centering
	\includegraphics[width=0.85\textwidth]{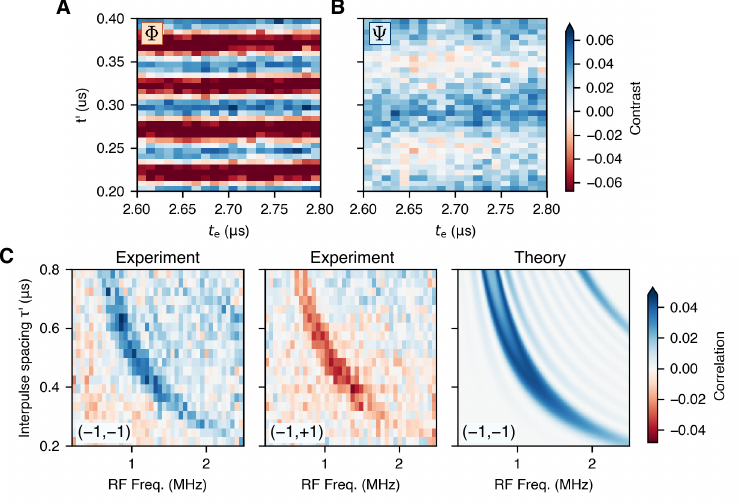}
	\caption{Bell state creation and correlations in the weak coupling regime, using the same NV pair used to collect the data in \cref{fig:entanglement_overview}F,G. The coupling strength is $J_{zz}\approx 2\pi\cdot183\,$kHz, and we use an entangling gate time of $t_\text{e}=2.732\,\mu$s. (A,B) Hahn echo experiments of $\ket{\Phi}$ (A) and $\ket{\Psi}$ (B) states using 10\,MHz TPPI. Residual 10\,MHz oscillations visible in the $\ket{\Psi}$ state indicate an admixture of single-quantum states. (C) Covariance noise spectroscopy across both RF frequency and interpulse spacing when addressing the same NV sublevels $(-1,-1)$ and opposite NV sublevels $(-1,+1)$, and comparison with theory (right). The sensitivity enhancement provided by the entanglement protocol enables multidimensional datasets to be acquired much more quickly.
	}
	\label{fig:SI_2Dcorr}
\end{figure*}

\section{Entanglement fidelity}

While full density matrix tomography is outside the scope of this work, here we use the TPPI experiments on the double-quantum coherences (e.g.\ \cref{fig:entanglement_overview}D) to establish lower bounds on our Bell state fidelity. The generic density matrix on the $0,1$ subspace may be written as
\begin{align}
\rho_\text{gen}=\frac{1}{2} \begin{pNiceMatrix}[cell-space-limits = 2pt,
rules/color=[gray]{0.9},rules/width=1pt]
\text{D}_0 & f_0 e^{-i \phi_f} & e_0 e^{-i \phi_e} &  a_0 e^{-i \phi_a} \\
f_0 e^{i \phi_f} & \text{C}_0 &  d_0 e^{-i \phi_d} & b_0 e^{-i \phi_b} \\
e_0 e^{i \phi_e} &  d_0 e^{i \phi_d} & \text{B}_0 &  c_0 e^{-i \phi_c} \\
 a_0 e^{i \phi_a} & b_0 e^{i \phi_b} & c_0 e^{i \phi_c} & \text{A}_0
\end{pNiceMatrix},
\end{align}
with the constraint that $\tfrac{1}{2}(A_0+B_0+C_0+D_0)=1$. Taking this to be the state after the entanglement operation, we seek to calculate the fidelity with respect to the Bell state density matrix, $\rho_\Phi=\frac{1}{2}(\ket{0,0}+i\ket{11})(\bra{0,0}+i\bra{11})$. Accounting for the probability $p_{\text{NV}^-}$ that each NV center is initialized in the correct (NV$^-$) charge state, we find:
\begin{align}
    F = p_{\text{NV}^-}^2\text{Tr}[\rho_\Phi\rho_\text{gen}]=\frac{p_{\text{NV}^-}^2}{2}\left[A_0/2+D_0/2+a_0\sin(\phi_a)\right]\geq p_{\text{NV}^-}^2\frac{a_0}{2}\left(1-\sin(\phi_a)\right),\label{eq:fidelity_bound}
\end{align}
where the final inequality holds because the presence of the density matrix coherences bounds the amplitudes of the corresponding state populations. To account for decoherence, we first define (see \cite{Dolde2013})
\begin{align}
    \sigma_{ab}&=\e^{-\chi_{\text{e},a}(t_\text{e})}+\e^{-\chi_{\text{e},b}(t_\text{e})} \\
    \delta_{ab}&=\e^{-\chi_{\text{e},a}(t_\text{e})}-\e^{-\chi_{\text{e},b}(t_\text{e})}\\
    \pi_{ab}&=\e^{-\chi_{\text{e},a}(t_\text{e})}\e^{-\chi_{\text{e},b}(t_\text{e})},
\end{align}
where $\chi_{\text{e},a,b}(t_\text{e})$ are the decoherence functions incurred during the Hahn echo entangling gates. Then the expected fidelity, adapting the expression shown in \cite{Dolde2013}, is 
\begin{align}
    F_\text{expected}=\frac{p_{\text{NV}^-}^2}{4}\left(1+\sigma_{ab}+\pi_{ab}\right).
\end{align}

We determine the values $a_0,\phi_a$ from the TPPI experiments, where we determine the signal from the expectation of the measurement operator, $M = 2\alpha_0\ket{00}\bra{00}+(\alpha_0+\alpha_1)(\ket{01}\bra{01}+\ket{10}\bra{10})+2\alpha_1\ket{11}\bra{11}$ (recall that $\alpha_i$ is the brightness of a single NV center in state $m_s=i$). Beginning with $\rho_\text{gen}$ as our entangled state, we apply the disentangling block $\overline{\mathcal{U}}_{\Psi}$ with each pulse's phase advanced by $\phi_\text{tppi}=\omega_\text{tppi}\tau$  (\cref{fig:entanglement_overview}D), and find the TPPI signal 
\begin{align}
S_\text{tppi}^\pm &= \text{Tr}[M\rho_\text{gen}^{\text{tppi},\pm}] \nonumber \\
& = (\alpha_0+\alpha_1) \pm \left( \alpha_0-\alpha_1\right) \left[\frac{\sigma_{ab}}{2}a_0\sin(2\phi_\text{tppi}-\phi_a) + \frac{\delta_{ab}}{2}d_0\sin(\phi_d)\right],
\end{align}
where the choice of $+$ or $-$ corresponds to the choice of phase in the final $\pi/2$ pulse in a contrast measurement.
The measured contrast signal is then
\begin{align}
    S_\text{tppi}^\text{cont.}\equiv \frac{S_\text{tppi}^+-S_\text{tppi}^-}{S_\text{tppi}^++S_\text{tppi}^-}=\left( \frac{\alpha_0-\alpha_1}{\alpha_0+\alpha_1}\right)\left[\frac{\sigma_{ab}}{2}a_0\sin(2\phi_\text{tppi}-\phi_a) + \frac{\delta_{ab}}{2}d_0\sin(\phi_d)\right].
\end{align}
Thus, after independently calibrating the NV center brightnesses $\alpha_0$ and $\alpha_1$, we may fit the amplitude and phase of the measured $2\phi_\text{tppi}$ TPPI oscillation to bound the fidelity. 

As an example, for the NV center pair whose data is shown in \cref{fig:entanglement_overview}D, the Hahn-echo coherence times of each NV center are approximately $T_{2,a} = 6\,\mu$s and $T_{2,b} = 12\,\mu$s. From our TPPI contrast measurement we find the fidelity bound (\cref{eq:fidelity_bound}) $F\geq 0.51 p_{\text{NV}^-}^2$, compared to the expected fidelity $F_\text{expect}=0.67 p_{\text{NV}^-}^2$. Our typical NV$^-$ initialization probability is $p_{\text{NV}^-}\approx0.75$, such that the true single-shot Bell state preparation fidelity is $F\gtrsim25\,\%$. Future experiments can boost the fidelity with charge state initialization (at the cost of increased experiment time), and incorporate more sophisticated pulse protocols designed to improve the entangling gate fidelity \cite{Dolde2014,Joas2024}.

\begin{figure*}[ht]
	\centering
	\includegraphics[width=0.95\textwidth]{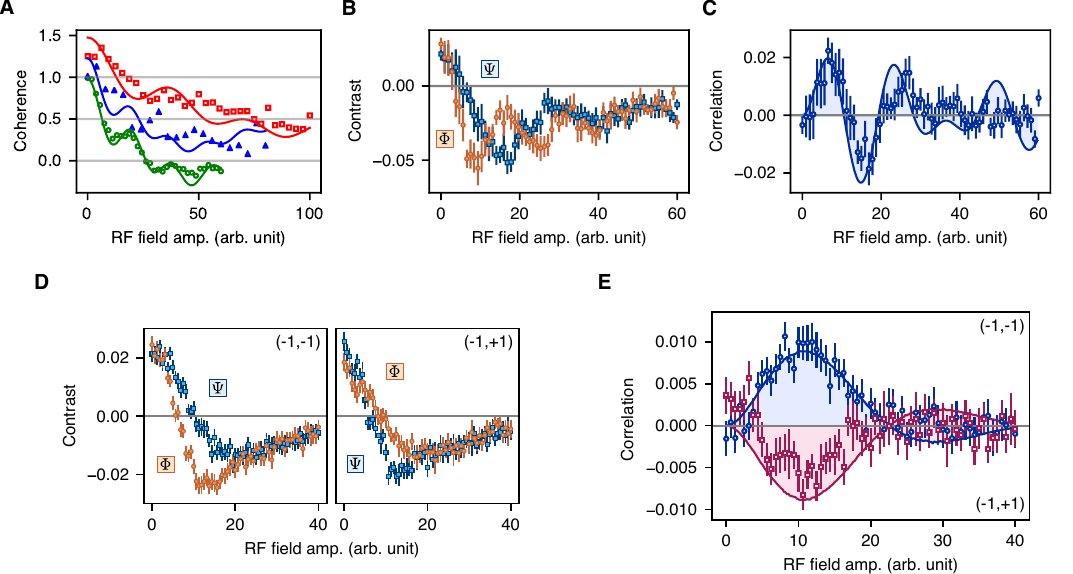}
	\caption{Calibrating expected noise correlations from the entangled NV centers used to acquire the data shown in \cref{fig:entanglement_overview}B,D. The coupling strength is approximately $J_{zz}=2\pi\cdot 250\,$kHz and we use an entangling gate time of $t_\text{e}=2\,\mu$s. (A) Conventional decoherence caused by an external 2 MHz RF test field with no NV-NV entanglement, using XY8 sensing of a pure tone (green circles), XY8 sensing of a field with added phase noise (blue triangles), and XY4 sensing of a field with added phase noise (red squares). Data are vertically offset for clarity. All fit lines derive from a single fitting model. (B) Entanglement-based super-resolution covariance magnetometry using XY8 sensing of a 2 MHz pure tone, measured while in the $\ket{\Psi}$ and $\ket{\Phi}$ states. (C) Correlation derived from the data in (B), and expected correlation (line) based on the calibration data in (A) with no free parameters except an overall amplitude scaling. Structure is caused by a significant difference in the noise amplitude experienced by the two NV centers, due to their different orientations. (D,E) Same as B,C, but the 2 MHz RF tone is phase-randomized with 1 MHz bandwidth white noise. Also shown are positive and negative correlations from addressing the same $(-1,-1)$ or opposite $(-1,+1)$ NV center transitions. 
	}
	\label{fig:SI_Calibrate}
\end{figure*}

\section{Calibration of expected correlations}

As described above (\cref{eq:cov_maximum}), the expected covariance amplitude from phase cycling is $\frac{1}{4}(\alpha_0-\alpha_1)^2 r_\text{ideal}$, which becomes $\frac{1}{8}(\alpha_0-\alpha_1)^2$ when saturating the noise sensing limit $r_\text{ideal} \leq 0.5$. For example, for the data shown in \cref{fig:overview}E, we find the single-NV brightness difference $\alpha_0-\alpha_1=0.48$ using spin-to-charge conversion, such that the maximum expected covariance is $\text{Cov} \approx 0.029$, in good agreement with the amplitude observed in \cref{fig:overview}E. The lines shown in \cref{fig:overview}E are derived from theory and independent calibration of the applied AC amplitude, with only an overall amplitude and decay profile added. We do not observe this decay in the single-NV data, and it is likely caused by the RF field becoming strong enough to have a non-negligible effect during the pulses, where we use a 56\,ns $\pi$ pulse duration.

To calibrate the expected correlation, we measure both NV centers using conventional XY4 and XY8 sensing of an external 2 MHz RF test tone (\cref{fig:SI_Calibrate}A). We fit a single model to three cases: XY8 sensing of a pure RF tone (the starting phase of the pure tone is random from one sequence to the next), XY8 sensing of an RF tone with 1 MHz phase modulation, and XY4 sensing of an RF tone with 1 MHz phase modulation. We then use Bell-pair sensing of a pure tone (\cref{fig:SI_Calibrate}B), and the resulting correlation measurement (\cref{fig:SI_Calibrate}C) corresponds well to the model predicted by the calibration from \cref{fig:SI_Calibrate}A. For the correlation data shown in \cref{fig:entanglement_overview}G, we expect the maximum achievable correlation to be $\tfrac{1}{2}\e^{-2\chi(t_\text{e})-\chi(N\tau)}(\alpha_0-\alpha_1)/(\alpha_0+\alpha_1)\approx0.06$, where the entanglement gate time is $t_\text{e}=2.732\,\mu$s and the sensing time is $N\tau=8\cdot500\,$ns. In practice we see a maximum correlation around $0.04$. The discrepancy is likely attributable to entanglement gate errors, where pulse durations are known to be detrimental \cite{Joas2024}, as well as initialization errors in the NV centers' charge states and spin states. 

\putbib[MultiQubitSensing]

\end{bibunit}

\end{document}